%% file: 10HzPRL.tex
\documentclass[10pt,aps,prl,twocolumn,superscriptaddress]{revtex4-2}

\usepackage{graphicx} 
\usepackage{dcolumn} 
\usepackage{bm}      
\usepackage{wasysym}
\usepackage[colorlinks=true,linkcolor=blue,citecolor=blue,urlcolor=blue]{hyperref}
\usepackage[utf8]{inputenc}

\begin{document}

\title{Continuous high-yield fast neutron generation with few-cycle laser pulses at 10 Hz for applications}

\author{L.~Stuhl}
\email[]{stuhl@atomki.hu}
\affiliation{National Laser-Initiated Transmutation Laboratory, University of Szeged, Szeged, 6720, Hungary}
\affiliation{HUN-REN Institute for Nuclear Research (HUN-REN ATOMKI), Debrecen, 4026, Hungary}
\affiliation{Center for Exotic Nuclear Studies, Institute for Basic Science, Daejeon 34126, Republic of Korea}

\author{P.~Varmazyar}
\affiliation{National Laser-Initiated Transmutation Laboratory, University of Szeged, Szeged, 6720, Hungary}

\author{Z.~Elekes}
\affiliation{HUN-REN Institute for Nuclear Research (HUN-REN ATOMKI), Debrecen, 4026, Hungary}
\affiliation{Institute of Physics, Faculty of Science and Technology, University of Debrecen, Debrecen, 4032, Hungary}

\author{Z.~Halász}
\affiliation{HUN-REN Institute for Nuclear Research (HUN-REN ATOMKI), Debrecen, 4026, Hungary}

\author{T.~Gilinger}
\affiliation{National Laser-Initiated Transmutation Laboratory, University of Szeged, Szeged, 6720, Hungary}

\author{M.~Füle}
\affiliation{National Laser-Initiated Transmutation Laboratory, University of Szeged, Szeged, 6720, Hungary}

\author{M.~Karnok}
\affiliation{National Laser-Initiated Transmutation Laboratory, University of Szeged, Szeged, 6720, Hungary}

\author{E.~Buzás}
\affiliation{National Laser-Initiated Transmutation Laboratory, University of Szeged, Szeged, 6720, Hungary}

\author{A.P.~Kovács}
\affiliation{National Laser-Initiated Transmutation Laboratory, University of Szeged, Szeged, 6720, Hungary}
\affiliation{Department of Optics and Quantum Electronics, University of Szeged, Szeged, 6720, Hungary}

\author{B.~Nagy}
\affiliation{National Laser-Initiated Transmutation Laboratory, University of Szeged, Szeged, 6720, Hungary}

\author{A.~Mohácsi}
\affiliation{National Laser-Initiated Transmutation Laboratory, University of Szeged, Szeged, 6720, Hungary}
\affiliation{ELI-ALPS, ELI-HU Non-Profit Ltd., Szeged, 6728, Hungary}

\author{B.~Bíró}
\affiliation{HUN-REN Institute for Nuclear Research (HUN-REN ATOMKI), Debrecen, 4026, Hungary}

\author{L.~Csedreki}
\affiliation{HUN-REN Institute for Nuclear Research (HUN-REN ATOMKI), Debrecen, 4026, Hungary}

\author{A.~Fenyvesi}
\affiliation{HUN-REN Institute for Nuclear Research (HUN-REN ATOMKI), Debrecen, 4026, Hungary}

\author{Zs.~Fülöp}
\affiliation{HUN-REN Institute for Nuclear Research (HUN-REN ATOMKI), Debrecen, 4026, Hungary}

\author{Z.~Korkulu}
\affiliation{HUN-REN Institute for Nuclear Research (HUN-REN ATOMKI), Debrecen, 4026, Hungary}
\affiliation{Center for Exotic Nuclear Studies, Institute for Basic Science, Daejeon 34126, Republic of Korea}

\author{I.~Kuti}
\affiliation{HUN-REN Institute for Nuclear Research (HUN-REN ATOMKI), Debrecen, 4026, Hungary}

\author{J.~Csontos}
\affiliation{ELI-ALPS, ELI-HU Non-Profit Ltd., Szeged, 6728, Hungary}
\author{P. P. Geetha}
\affiliation{ELI-ALPS, ELI-HU Non-Profit Ltd., Szeged, 6728, Hungary}
\author{Sz.~Tóth}
\affiliation{ELI-ALPS, ELI-HU Non-Profit Ltd., Szeged, 6728, Hungary}

\author{G.~Szabó}
\affiliation{Department of Optics and Quantum Electronics, University of Szeged, Szeged, 6720, Hungary}
\affiliation{ELI-ALPS, ELI-HU Non-Profit Ltd., Szeged, 6728, Hungary}

\author{K.~Osvay}
\email[]{osvay@physx.u-szeged.hu}
\affiliation{National Laser-Initiated Transmutation Laboratory, University of Szeged, Szeged, 6720, Hungary}
\affiliation{Department of Optics and Quantum Electronics, University of Szeged, Szeged, 6720, Hungary}


\date{\today}

\begin{abstract}
We present a laser-based neutron source that produces \(1.8 \times 10^5\,\text{neutrons/s}\) with a conversion rate of \(7.8 \times 10^5\,\text{neutrons/J}\). Laser pulses of 12~fs and 23~mJ were focused onto a \(430\text{-nm}\)-thick heavy water liquid sheet at a 10~Hz repetition rate. The resulting peak intensity of \(4 \times 10^{18}\,\text{W/cm}^2\) accelerated deuterium ions from the target rear side to a kinetic energy of 1~MeV. This deuteron beam induced \(^{2}\text{H}(d,n)^{3}\text{He}\) fusion reactions in a deuterated polyethylene target, producing fast neutrons. The neutron yield was measured using two independent detection systems: the LILITH time-of-flight spectrometer, consisting of eight plastic scintillators covering nearly \(180^\circ\), and a calibrated bubble detector spectrometer. The neutron yield per laser shot is 35 times higher than that recently achieved by lasers with comparable pulse energies, while the conversion rate is the highest ever achieved by continuously operating, sub-100~fs lasers. The generated neutrons are emitted from an area of \(0.65~\text{cm}^2\) corresponding to the deuteron beam spot on the catcher. Their angular distribution is peaked in forward and backward directions in agreement with the literature data on the angular distribution of \(^{2}\text{H}(d,n)^{3}\text{He}\) reaction. The system operated continuously for several hours per day with an unprecedented stability of \(5\%\).

\end{abstract}

\maketitle
Laser-based neutron generation has attracted significant scientific interest over the past decades for two main reasons. First, the fast neutron pulses produced are several orders of magnitude shorter in duration than those generated by conventional spallation sources. Therefore, the resulting ultra-high peak neutron flux is beneficial, e.g., for fusion materials research and neutron resonance spectroscopy~\cite{Perkins2000, Yogo2023advances}. Second, the associated radiation safety requirements are comparatively less demanding. For laser-driven systems, radiation protection is primarily required for the source of the neutrons, typically around a meter-scale vacuum chamber, while the laser system itself does not require additional shielding~\cite{Brenner2016, Roth2017, Mori2023, Zimmer2024, Mirfayzi2025, Canova2025}.

Two laser-based neutron generation schemes have been widely studied. The first is the pitcher-catcher configuration, in which ions are accelerated from an initial target (the pitcher) and subsequently interact with a secondary target (the catcher) to produce neutrons~\cite{Norreys1998, Disdier1999}. To eliminate the need for two targets, direct neutron generation from a single bulk target has also been investigated~\cite{Ditmire1999, Kitagawa2011, Wang2013}. A comparative study~\cite{Willingale2011} suggested that low-energy laser systems are suitable for direct neutron generation from bulk targets, whereas high-energy, high-peak intensity laser systems provide the highest neutron yields using the pitcher-catcher scheme. The highest neutron yields to date, exceeding \(10^{11}\, \text{neutrons/shot}\), have been achieved using hundred-joule picosecond-duration lasers operated in single-shot mode~\cite{Guenther2022, Yogo2023}.

Most real-world applications require a continuously operational neutron source driven by table-top, short-pulse lasers with watt-class average power. A critical challenge in this regime is the development of (pitcher) target systems that can precisely renew the target material at the correct position and orientation before each laser pulse~\cite{Prencipe2017}. Target systems with thin foils limits the achievable repetition rate to about 1 Hz \cite{Osvay2024_1Hz} and the number of shots to a few hundred \cite{Lelievre2024}. The latter experiment using laser pulses with an energy of few joules demonstrated the so far record laser-to-neutron conversion rate of \(1.56 \times 10^5\, \text{neutrons/J}\). An alternative approach is to use lower-energy lasers operating at higher repetition rates \cite{Osvay2024towards, treffert2021towards}. Liquid jets and flat sheet targets have shown great promise for high-repetition-rate neutron sources. Early experiments with 18 mJ laser pulses at \(500\,\text{Hz}\)~\cite{Hah2016, Hah2018} achieved a neutron yield of \(2 \times 10^5\,\text{neutrons/s}\), where neutrons were emitted directly from the laser-induced heavy-water plasma. The conversion rate is ~\(2.2\times 10^4\,\text{neutrons/J}\), with a nearly isotropic neutron distribution, and an estimated flux below \(10^2\,\text{neutrons/cm}^2/\text{s}\). In a recent follow-up experiment, half of that performance was demonstrated at 1 kHz ~\cite{Knight2024}  Recently, flat jet systems have been further advanced for high-repetition-rate proton and deuteron acceleration, targeting applications in the pitcher-catcher scheme~\cite{Treffert2022, Fule2024}.

In this Letter, we present a novel femtosecond-laser–based neutron source with a yield of \(1.8 \times 10^5\,\text{neutrons/s}\), the first of its kind, capable of operating for several hours per day. The record laser-to-neutron conversion rate of \(7.8 \times 10^5\, \text{neutrons/J}\) was achieved by optimized acceleration of deuteron ions from a heavy water liquid sheet. We provide a detailed kinematic characterization of neutrons from \(^{2}\text{H}(d,n)^{3}\text{He}\) fusion, including energy and angular distributions, confirming negligible tritium-related neutron production.

An off-axis parabola was used to focus the p-polarized pulses from the SEA laser \cite{Toth2020} of ELI-ALPS onto a heavy-water (D\textsubscript{2}O) liquid sheet at a \(45^\circ\) incidence angle. The energy and duration of the laser pulses, measured at the position of the target, were \(23\,\text{mJ}\) and \(12\,\text{fs}\), respectively. The focal spot, measuring \(3 \times 3\,\mu\text{m}^2\), contained 36\% of the total pulse energy, producing a peak intensity of \(4 \times 10^{18}\,\text{W/cm}^2\) on target. Accelerated deuterium ions were detected using two Thomson Parabola Spectrometers (TPS): one positioned in the forward direction (detecting ions from the rear side of the target) and one in the backward direction (detecting ions from the front side), as shown in Fig.~1(a). The ion tracks, deflected within each TPS, were recorded using CCD cameras coupled to a phosphor screen and a microchannel plate (MCP). The assembly was calibrated for deuterium ions prior to the experiment, as described in \cite{Osvay2024towards}.

\begin{figure}
\includegraphics[width=85mm,angle=0]{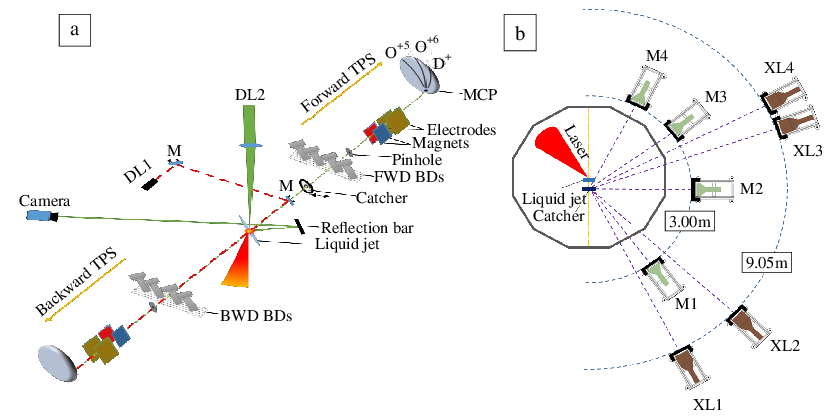}
\caption{(a) Monitoring setup for deuterons and heavy water jet alignment. (b) Schematic view of the target chamber and LILITH detectors.}
\end{figure}

The flat sheet of heavy water was produced by two colliding liquid jets generated by micro-nozzles with a diameter of \(18\,\mu\text{m}\). The sheet thickness is determined by the nozzle size, the flow rate, and the distance from the jet collision point~\cite{Fule2024}. To avoid ablating the liquid catcher's orifice and destabilizing the jet, we chose  a position on the sheet resulting in a thickness of \(430\,\text{nm}\). The sheet was aligned perpendicular to the TPS axes using a red diode laser (DL1 in Fig.~1(a)). The sheet position was continuously monitored using a green diode laser (DL2) focused on its surface. A camera outside the target chamber observed the green spot reflected from the sheet surface. If the green spot shifted, the jet position was manually adjusted to restore the sheet.

The catcher target was manufactured from deuterated polyethylene (dPE) powder compressed into a tablet with a diameter of \(30\,\text{mm}\) and thickness of \(1\,\text{mm}\). A \(1\,\text{mm}\)-diameter hole drilled at the center of the tablet enabled continuous monitoring of the deuteron beam by the forward TPS. The tablet was mounted on a translation stage, allowing it to move into or out of the deuteron beam path.

For event-by-event characterization of neutrons, the LILITH time-of-flight (ToF) spectrometer, consisting of eight organic scintillators, was developed. Four of these, LILITH-M, are cylindrical EJ-230 fast plastic scintillators (\(\diameter150\,\text{mm} \times 25\,\text{mm}\)) coupled to Hamamatsu R2083 photomultiplier tubes (PMTs) via light guides. The other four detectors, LILITH-XL, are cylindrical EJ-309 liquid scintillators (\(\diameter127\,\text{mm} \times 127\,\text{mm}\)) coupled to ET Enterprises 9390B PMTs. The M and XL detectors were placed approximately \(3\,\text{m}\) and \(9\,\text{m}\), respectively, from the catcher target. The positions and angles of all detectors relative to the catcher and the deuteron beam were determined photogeometrically (Fig.~1(b)): LILITH-M detectors at \(37^\circ\) (M1), \(88.8^\circ\) (M2), \(126.2^\circ\) (M3), \(151.4^\circ\) (M4), and LILITH-XL detectors at \(34.6^\circ\) (XL1), \(113^\circ\) (XL2), \(118.4^\circ\) (XL3), \(43.6^\circ\) (XL4).

All LILITH detectors were shielded with \(5\,\text{cm}\) of lead against X-rays and gamma rays. The light output of the LILITH detectors was calibrated using \({}^{137}\text{Cs}\), \({}^{60}\text{Co}\), and \({}^{22}\text{Na}\) gamma sources, while their neutron detection efficiencies were determined with a plutonium-beryllium (PuBe) source. Waveforms were digitized over \(1400\,\text{ns}\) and analyzed offline. The details of the calibration and data processing are discussed in~\cite{Stuhl2020, Osvay2024_1Hz}. In this geometry, the energy resolution is around \(1\%\) for M and \(1.5\%\) for XL detectors at \(2.5\,\text{MeV}\).

In parallel, a bubble neutron detector spectrometer (BDS), insensitive to electromagnetic pulses and gamma radiation, provided a secondary measurement. The detectors were placed inside the target chamber using vacuum-tight aluminum tubes (Fig.~1(a)) along the axis of the deuteron beam. As factory calibration is reliable for \(\geq 100\) bubbles, the individual detectors were calibrated for low counts using a PuBe source~\cite{Osvay2024towards}.

Several measurement runs were performed during the experimental campaign, varying the laser pulse energy, duration, and repetition rate, both with and without the catcher. In addition to neutron generation, the experiment was designed to irradiate biological samples~\cite{Osvay2024towards}. Each day, over 70,000 shots were delivered, with intermittent interruptions due to liquid nitrogen refills to maintain the vacuum, exchange of the catcher, biological sample insertions, or temporary remote connection losses. This report focuses on data from the last day, which included two long, continuous sessions. The LILITH data correspond to the second session, the longest of the day.

\begin{figure}
\includegraphics[width=85mm,angle=0]{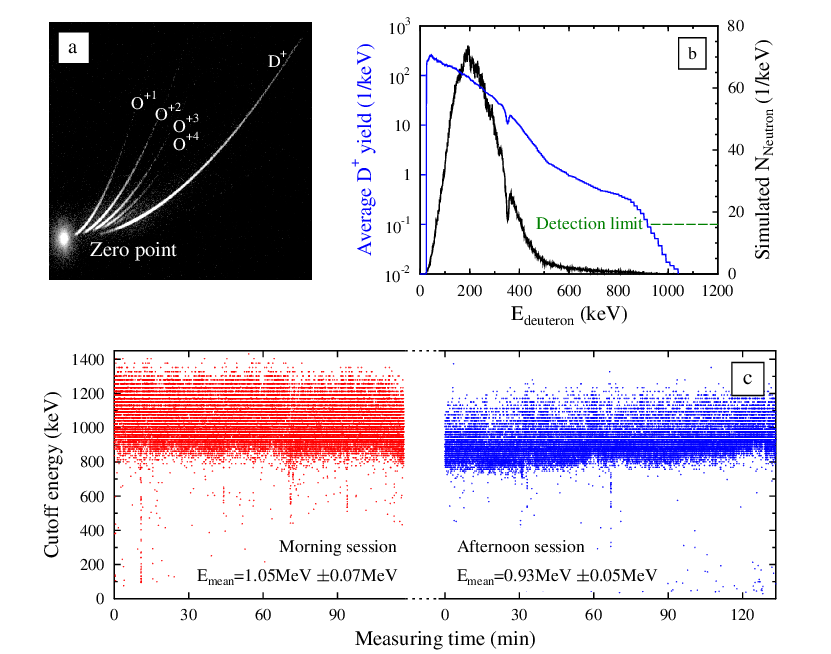}
\caption{(a) Typical image captured by the forward TPS camera. (b) Average deuteron energy spectrum (blue) from the afternoon run, and the simulated neutron yield (black) calculated using the \(^{2}\text{H}(d,n)^{3}\text{He}\) fusion cross section. (c) Cutoff energy distribution of deuterons accelerated by all laser shots of the two longest runs of the day. }
\end{figure}

Throughout the experiment, the ion spectra were recorded for each laser shot. A typical ion track is shown in Fig.~2(a), where \(\text{D}^+\) dominates, while the proton line from \(0.25\%\) water contamination is barely visible. To determine the highest cutoff energy of the deuterons, the pulse duration was stretched using the acousto-optical programmable filter of the laser, in a procedure similar to that described in \cite{Ziegler2021}. For the heavy water sheet with a thickness of 430 nm, the optimal pulse duration was found to be about 200 fs \cite{Osvay2024AAC}. Considering the high temporal purity of the pulses, the acceleration mechanism is target-normal-sheath acceleration (TNSA), with slightly expanding plasma formation at the front side of the sheet \cite{Schreiber2014, Ziegler2021}. The blue line in Fig.~2(b) represents the average deuteron energy spectrum for the longest run. The black line, simulated with \textsc{Geant4}~\cite{geant4} using the average deuteron spectrum, the \(^2\mathrm{H}(d,n)^3\mathrm{He}\) reaction cross section, and the catcher parameters. This indicates that 255~keV deuterons contribute most significantly to neutron generation.

To demonstrate the stability of the laser plasma acceleration  during continuous operation, Fig.~2(c) shows the cutoff energies of the measured deuteron spectra. As is known, a liquid sheet target is inherently unstable during laser operation~\cite{CaoFrontiers2023, Fule2024, Peng_HPLSE_2024}, exhibiting stochastic and periodic rotation around its steady position. A TNSA ion beam exhibits a strong energy gradient, with the fastest particles propagating near the center of the beam and the slower particles at the edges~\cite{Zeil_2010}. Due to target rotation, different parts of the ion beam propagate through the pinhole of the TPSs. Therefore, the determined \(10\%\) uncertainty is a conservative estimate and results from a combination of laser fluctuations, target position, and, especially, target angle instabilities.

\begin{figure}
\includegraphics[width=85mm,angle=0]{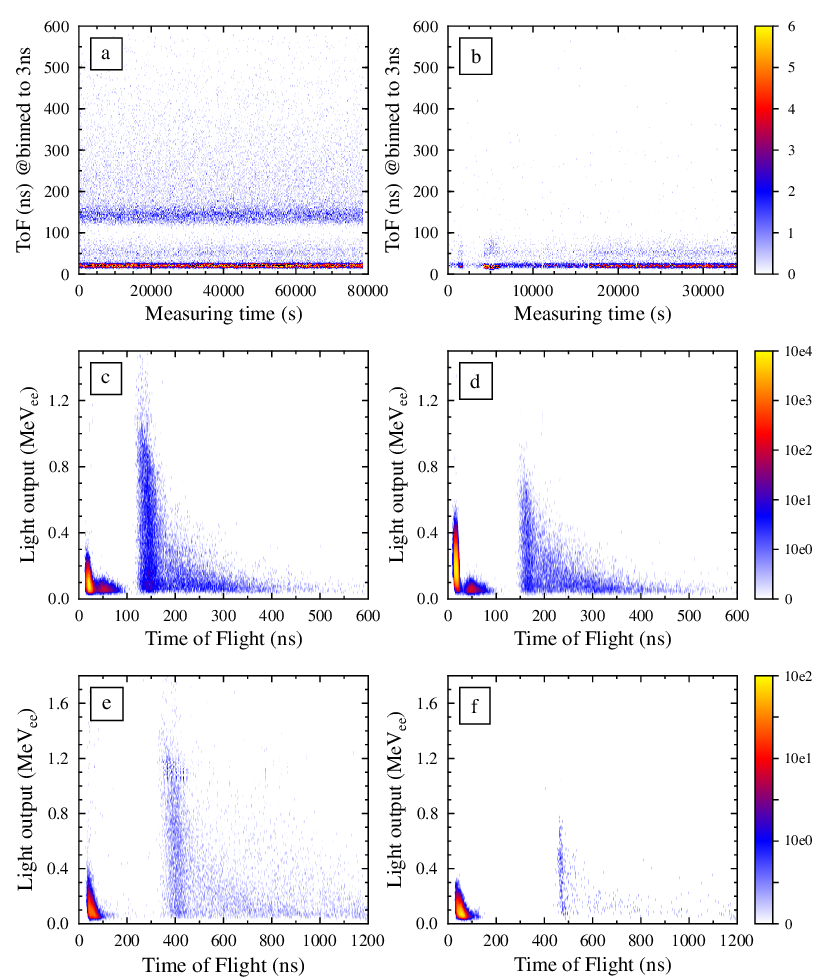}
\caption{(a) Measured ToF data as a function of time for the forward-positioned LILITH M1 detector during a two-hour period with the catcher in the deuteron beam. (b) ToF data for a one-hour period with the catcher removed. (c) Response of LILITH M1 to fusion neutrons and X-ray/gamma-ray bursts as a function of ToF between the PD and detector. Panels (d), (e), and (f) show corresponding responses for the LILITH M2, LILITH XL1, and LILITH XL3 detectors, respectively.}
\end{figure}

The ToF information was extracted by measuring the time difference between a reference signal from a photodiode (PD) and the detector signal in an event-by-event analysis. A strong prompt signal from a burst of laser-driven X-rays and gamma rays served as the time reference for ToF calibration. Figure~3(a) shows the ToF data recorded by the forward-oriented LILITH M1 detector for the longest run comprising 79,808 laser shots. During this run, the catcher was present, resulting in a significant number of events with longer ToF. In contrast, Fig.~3(b) displays data from a background run of 34,304 shots, where the catcher was absent, with a few detected events beyond the 100~ns ToF range. This comparison, consistent with previous findings~\cite{higginson2024}, demonstrates that the primary source of X-ray and gamma radiation detected by the LILITH system was the pitcher target. Furthermore, the uniformity of the ToF distribution over time is noteworthy.

Calibrated light output versus offset-corrected ToF for X-rays, gamma rays (\(<100~\text{ns}\)) and neutrons demonstrate the operation of four LILITH detectors (Fig.~3(c)–(f)): forward-oriented M1 (Fig.~3(c)), perpendicular M2 (Fig.~3(d)), forward XL1 (Fig.~3(e)), and backward XL3 (Fig.~3(f)). Sharp peaks beginning at 10~ns (for M detectors) and 30~ns (for XL detectors) correspond to photons. Events with ToF beyond 110~ns (M) and 330~ns (XL) are attributed to neutrons. The neutron distributions exhibit a sharp peak (unscattered fusion neutrons) followed by a broader distribution, caused by scattering within the experimental environment. The ToF peaks shift to higher values with increasing laboratory angle, while the peak light output decreases, indicating a clear directional dependence. Notably, the XL detector data reveal no evidence of 14.1~MeV neutrons from the \(^{3}\text{H}(d,n)^{4}\text{He}\) reaction.

The timing uncertainty of deuterons traveling from the pitcher to the catcher depends on their kinetic energy. Since event-by-event corrections were not feasible, an offline timing adjustment of 29~ns  was applied, corresponding to the ToF of 255~keV deuterons over a 14.3~cm path, as derived from the simulated neutron yield (Fig.~2(b)). Corrected neutron energies were then calculated using this adjustment and the measured ToF path lengths.

\begin{figure}
\includegraphics[width=85mm,angle=0]{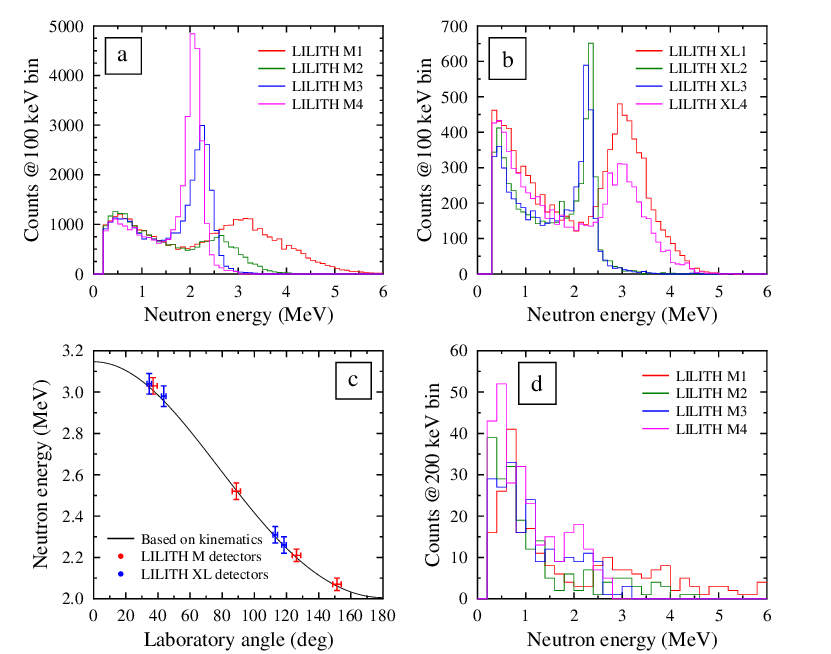}
\caption{(a) Energy distribution of neutrons detected by the four LILITH M detectors. (b) Energy distribution of neutrons detected by the LILITH XL detectors. (c) Directional dependence of deduced energy centroids compared to calculated distributions for \(^{2}\text{H}(d,n)^{3}\text{He}\) reactions with \(E_k(\text{deuteron}) = 255\,\text{keV}\) (see Fig.~2(b)). (d) Energy distribution recorded by the LILITH M detectors without the catcher in the deuteron beam.}
\end{figure} 

Neutron yields were determined from the ToF distributions by integrating events within the 110–500~ns (M) and 330–1400~ns (XL) ranges. In the longest run (79,808 shots), over 130{,}000 and 30{,}000 neutrons were detected in the M and XL detectors, respectively. The average neutron yield for the entire run was \(17,960 \pm 580\,\text{neutrons/shot}\), while the peak yield, calculated over a 5-minute interval, reached \(21,000 \pm 760\,\text{neutrons/shot}\). These yields represent the total production averaged over the \(4\pi\) solid angle, taking into account the anisotropic neutron distribution. The total uncertainty comprises statistical and systematic components; the latter include uncertainties in neutron event identification (1\%), interaction point and angle (3--6\%), and variations in intrinsic efficiencies and light output threshold determination (5\%).

Figure~4(a) shows neutron energy spectra from the LILITH M detectors, while Fig.~4(b) displays spectra from the XL detectors. Lower-energy events correspond to scattered neutrons, while peaks represent unscattered neutrons. Approximately \(56 \pm 5\%\) of the events in the LILITH M detectors were unscattered. The peaks were fitted, and the mean values of the distributions were compared with the expected mean values at specific laboratory recoil angles, calculated using relativistic kinematics for nuclear reactions. As shown in Fig.~4(c), the experimental results are in good agreement with calculations~\cite{catkin2019}, providing direct evidence that the detected neutrons primarily originate from the reaction \(^{2}\text{H}(d,n)^{3}\text{He}\).

\begin{figure}
\includegraphics[width=85mm,angle=0]{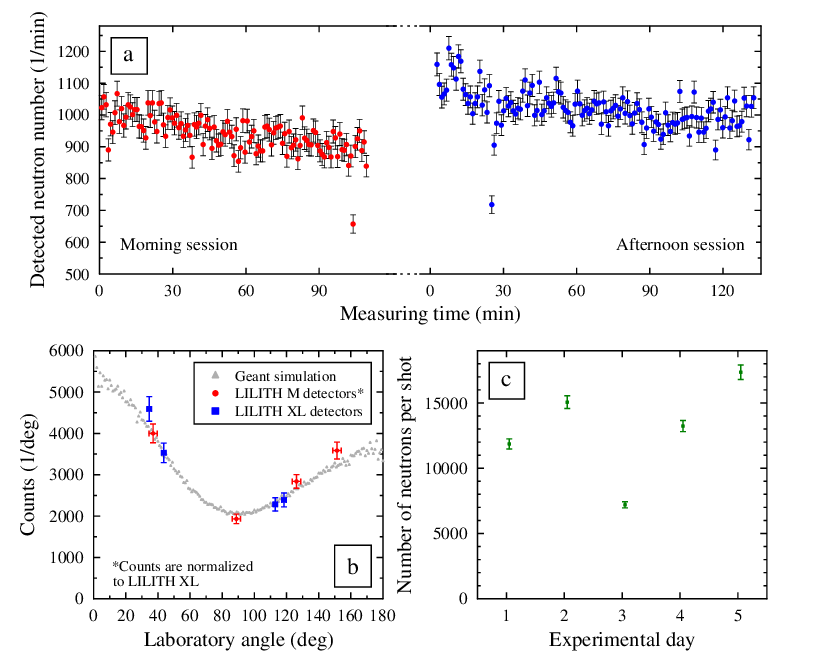}
\caption{(a) Total neutron events per minute recorded by LILITH M detectors over the entire day. (b) Simulated angular distribution, based on the average deuteron spectrum from the longest run, compared to neutron counts at various laboratory angles, normalized to the solid angle of the LILITH XL detectors. LILITH M counts are also normalized to those of the XL detectors. (c) Variation of the determined average neutron yield per shot across the five experimental days.}
\end{figure}

In order to investigate the amount of neutrons generated directly in the heavy water liquid sheet, as in~\cite{Knight2024, Karsch2003}, a 58-minute run was performed without catcher. The neutron energy spectra (Fig.~4(d)) from the LILITH M detectors reveal a very low sub-peak around 2~MeV. By normalizing the background measurement to the longest reported run, it was calculated that only \(1.59 \pm 0.07\%\) of the detected neutrons originated from the pitcher target. This provides clear experimental evidence that the pitcher-catcher scheme significantly outperforms direct neutron generation.

The stability of the neutron generation was analyzed by calculating the total neutron count for 1-minute intervals (Fig.~5(a)). The RMS stability of the neutron yield was \(5\%\) during the longest run and \(7\%\) over the entire day. In particular, the neutron yield exhibited a continuous decline in both periods, while the deuteron cutoff energies remained stable over time (see Fig.~2(c)). This behavior was attributed to the surface degradation (burning) of the otherwise static dPE catcher. Remarkably, the stability of the neutron source appears to be significantly higher than that of the deuteron cutoff energies, indicating that the real energy stability of the deuteron accelerator is greater than the cutoff stability shown in Fig.~2(c).

The measured angular distribution of \(^{2}\text{H}(d,n)^{3}\text{He}\) fusion neutrons is consistent with \textsc{Geant4} simulations based on the average deuteron spectrum displayed in Fig.~2(b). The distribution is slightly more forward oriented than previously observed~\cite{Osvay2024_1Hz}, corresponding to the larger average kinetic energy of the deuterons in the present study. This behavior also meets the prediction from earlier nuclear physics studies of the reaction \(^{2}\text{H}(d,n)^{3}\text{He}\)~\cite{Brown1990,Ying1973}, and the first analysis of beam-like neutron production with lasers \cite{Kar_2016}.

Figure~5(c) shows the neutron yield per shot for each day of the experiment, measured with the LILITH spectrometer. The BDS values are systematically four times higher. This is consistent with the report of other study \cite{Knight2024}, although the discrepancy here is two orders of magnitude smaller, probably due to the calibration of our BDS. However, to remain conservative, we report only the ToF-based yields. The laser-driven neutron source demonstrated a day-to-day stability within 25\% of the average yield, $1.4 \times 10^5$~neutrons/s. The weighted mean from daily shot counts confirms stable performance over 20 hours of operation.

The size of the neutron source is inherently identical to that of the deuteron beam reaching the catcher. The latter imprinted on the front surface of the catcher tablet showed a diameter of 9.1 mm. A sample placed 1 mm behind the catcher receives an average neutron flux of \(1.2 \times 10^5\,\text{neutrons/cm}^2/\text{s}\), which is sufficient to initiate radiobiological studies~\cite{Osvay2024towards}.

In conclusion, we report a new laser-based neutron source that reaches a yield of \(1.8 \times 10^5\, \text{neutrons/s}\) using only \(23\,\text{mJ}\) laser pulses at a repetition rate of \(10\,\text{Hz}\). The generated fast neutrons were measured with two independent detection systems, confirming the reliability of the conservative results. The measured neutron yield per laser shot was 35 times higher than that recently achieved by a similar laser system but with a single target configuration. The laser-to-neutron conversion rate is the highest ever demonstrated using femtosecond laser systems. The system showed continuous operation at a repetition rate of \(10\,\text{Hz}\) for several hours per day with an unprecedented stability of \(5\%\). The daily performance was so convincing that ELI-ERIC offers this laser neutron source to its user community.

\begin{acknowledgments}
This work has been supported by the National Research, Development, and Innovation Office through the National Laboratory Program (contracts NKFIH-877-2/2020 and NKFIH-476-4/2021) and by the European Union (MULTISCAN 3D, H2020-SU-SEC-2020 RIA, grant \#101020100). The ELI-ALPS project (GINOP-2.3.6-15-2015-00001) is supported by the European Union and co-financed by the European Regional Development Fund. ZK and LS acknowledge the support of the Institute for Basic Science (IBS-R031-D1).
\end{acknowledgments}

\input{10HzPRL.bbl}

\end{document}

%% file: 10HzPRL.bbl
%

%% file: 10HzPRL.bbl
\begin{thebibliography}{41}%
\makeatletter
\providecommand \@ifxundefined [1]{%
 \@ifx{#1\undefined}
}%
\providecommand \@ifnum [1]{%
 \ifnum #1\expandafter \@firstoftwo
 \else \expandafter \@secondoftwo
 \fi
}%
\providecommand \@ifx [1]{%
 \ifx #1\expandafter \@firstoftwo
 \else \expandafter \@secondoftwo
 \fi
}%
\providecommand \natexlab [1]{#1}%
\providecommand \enquote  [1]{``#1''}%
\providecommand \bibnamefont  [1]{#1}%
\providecommand \bibfnamefont [1]{#1}%
\providecommand \citenamefont [1]{#1}%
\providecommand \href@noop [0]{\@secondoftwo}%
\providecommand \href [0]{\begingroup \@sanitize@url \@href}%
\providecommand \@href[1]{\@@startlink{#1}\@@href}%
\providecommand \@@href[1]{\endgroup#1\@@endlink}%
\providecommand \@sanitize@url [0]{\catcode `\\12\catcode `\$12\catcode
  `\&12\catcode `\#12\catcode `\^12\catcode `\_12\catcode `\%12\relax}%
\providecommand \@@startlink[1]{}%
\providecommand \@@endlink[0]{}%
\providecommand \url  [0]{\begingroup\@sanitize@url \@url }%
\providecommand \@url [1]{\endgroup\@href {#1}{\urlprefix }}%
\providecommand \urlprefix  [0]{URL }%
\providecommand \Eprint [0]{\href }%
\providecommand \doibase [0]{https://doi.org/}%
\providecommand \selectlanguage [0]{\@gobble}%
\providecommand \bibinfo  [0]{\@secondoftwo}%
\providecommand \bibfield  [0]{\@secondoftwo}%
\providecommand \translation [1]{[#1]}%
\providecommand \BibitemOpen [0]{}%
\providecommand \bibitemStop [0]{}%
\providecommand \bibitemNoStop [0]{.\EOS\space}%
\providecommand \EOS [0]{\spacefactor3000\relax}%
\providecommand \BibitemShut  [1]{\csname bibitem#1\endcsname}%
\let\auto@bib@innerbib\@empty
\bibitem [{\citenamefont {Perkins}\ \emph {et~al.}(2000)\citenamefont
  {Perkins}, \citenamefont {Logan}, \citenamefont {Rosen}, \citenamefont
  {Perry}, \citenamefont {de~la Rubia}, \citenamefont {Ghoniem}, \citenamefont
  {Ditmire}, \citenamefont {Springer},\ and\ \citenamefont
  {Wilks}}]{Perkins2000}%
  \BibitemOpen
  \bibfield  {author} {\bibinfo {author} {\bibfnamefont {L.}~\bibnamefont
  {Perkins}}, \bibinfo {author} {\bibfnamefont {B.}~\bibnamefont {Logan}},
  \bibinfo {author} {\bibfnamefont {M.}~\bibnamefont {Rosen}}, \bibinfo
  {author} {\bibfnamefont {M.}~\bibnamefont {Perry}}, \bibinfo {author}
  {\bibfnamefont {T.~D.}\ \bibnamefont {de~la Rubia}}, \bibinfo {author}
  {\bibfnamefont {N.}~\bibnamefont {Ghoniem}}, \bibinfo {author} {\bibfnamefont
  {T.}~\bibnamefont {Ditmire}}, \bibinfo {author} {\bibfnamefont
  {P.}~\bibnamefont {Springer}},\ and\ \bibinfo {author} {\bibfnamefont
  {S.}~\bibnamefont {Wilks}},\ }\bibfield  {title} {\bibinfo {title} {The
  investigation of high intensity laser driven micro neutron sources for fusion
  materials research at high fluence},\ }\href
  {https://doi.org/10.1088/0029-5515/40/1/301} {\bibfield  {journal} {\bibinfo
  {journal} {Nuclear Fusion}\ }\textbf {\bibinfo {volume} {40}},\ \bibinfo
  {pages} {1} (\bibinfo {year} {2000})}\BibitemShut {NoStop}%
\bibitem [{\citenamefont {Yogo}\ \emph
  {et~al.}(2023{\natexlab{a}})\citenamefont {Yogo}, \citenamefont {Arikawa},
  \citenamefont {Abe}, \citenamefont {Mirfayzi}, \citenamefont {Hayakawa},
  \citenamefont {Mima},\ and\ \citenamefont {Kodama}}]{Yogo2023advances}%
  \BibitemOpen
  \bibfield  {author} {\bibinfo {author} {\bibfnamefont {A.}~\bibnamefont
  {Yogo}}, \bibinfo {author} {\bibfnamefont {Y.}~\bibnamefont {Arikawa}},
  \bibinfo {author} {\bibfnamefont {Y.}~\bibnamefont {Abe}}, \bibinfo {author}
  {\bibfnamefont {S.~R.}\ \bibnamefont {Mirfayzi}}, \bibinfo {author}
  {\bibfnamefont {T.}~\bibnamefont {Hayakawa}}, \bibinfo {author}
  {\bibfnamefont {K.}~\bibnamefont {Mima}},\ and\ \bibinfo {author}
  {\bibfnamefont {R.}~\bibnamefont {Kodama}},\ }\bibfield  {title} {\bibinfo
  {title} {Advances in laser-driven neutron sources and applications},\ }\href
  {https://doi.org/10.1140/epja/s10050-023-01083-8} {\bibfield  {journal}
  {\bibinfo  {journal} {The European Physical Journal A}\ }\textbf {\bibinfo
  {volume} {59}},\ \bibinfo {pages} {191} (\bibinfo {year}
  {2023}{\natexlab{a}})}\BibitemShut {NoStop}%
\bibitem [{\citenamefont {Brenner}\ \emph {et~al.}(2016)\citenamefont
  {Brenner}, \citenamefont {Mirfayzi}, \citenamefont {Rusby}, \citenamefont
  {Armstrong}, \citenamefont {Alejo}, \citenamefont {Wilson}, \citenamefont
  {Clarke}, \citenamefont {Ahmed}, \citenamefont {Butler}, \citenamefont
  {Haddock} \emph {et~al.}}]{Brenner2016}%
  \BibitemOpen
  \bibfield  {author} {\bibinfo {author} {\bibfnamefont {C.~M.}\ \bibnamefont
  {Brenner}}, \bibinfo {author} {\bibfnamefont {S.~R.}\ \bibnamefont
  {Mirfayzi}}, \bibinfo {author} {\bibfnamefont {D.~R.}\ \bibnamefont {Rusby}},
  \bibinfo {author} {\bibfnamefont {C.}~\bibnamefont {Armstrong}}, \bibinfo
  {author} {\bibfnamefont {A.}~\bibnamefont {Alejo}}, \bibinfo {author}
  {\bibfnamefont {L.~A.}\ \bibnamefont {Wilson}}, \bibinfo {author}
  {\bibfnamefont {R.}~\bibnamefont {Clarke}}, \bibinfo {author} {\bibfnamefont
  {H.}~\bibnamefont {Ahmed}}, \bibinfo {author} {\bibfnamefont {N.~M.~H.}\
  \bibnamefont {Butler}}, \bibinfo {author} {\bibfnamefont {D.}~\bibnamefont
  {Haddock}}, \emph {et~al.},\ }\bibfield  {title} {\bibinfo {title}
  {Laser-driven x-ray and neutron source development for industrial
  applications of plasma accelerators},\ }\href
  {https://doi.org/10.1088/0741-3335/58/1/014039} {\bibfield  {journal}
  {\bibinfo  {journal} {Plasma Physics and Controlled Fusion}\ }\textbf
  {\bibinfo {volume} {58}},\ \bibinfo {pages} {014039} (\bibinfo {year}
  {2016})}\BibitemShut {NoStop}%
\bibitem [{\citenamefont {Roth}\ \emph {et~al.}(2017)\citenamefont {Roth},
  \citenamefont {Vogel}, \citenamefont {Bourke}, \citenamefont {Fernandez},
  \citenamefont {Mocko} \emph {et~al.}}]{Roth2017}%
  \BibitemOpen
  \bibfield  {author} {\bibinfo {author} {\bibfnamefont {M.}~\bibnamefont
  {Roth}}, \bibinfo {author} {\bibfnamefont {S.}~\bibnamefont {Vogel}},
  \bibinfo {author} {\bibfnamefont {M.}~\bibnamefont {Bourke}}, \bibinfo
  {author} {\bibfnamefont {J.}~\bibnamefont {Fernandez}}, \bibinfo {author}
  {\bibfnamefont {M.}~\bibnamefont {Mocko}}, \emph {et~al.},\ }\href@noop {}
  {\bibinfo {title} {Assessment of laser-driven pulsed neutron sources for
  poolside neutron-based advanced {NDE}-{A} pathway to {LANSCE}-like
  characterization at {INL}}},\ \bibinfo {howpublished} {Los Alamos National
  Lab. Los Alamos, NM, U.S.A., report No. LA-UR-17-23190} (\bibinfo {year}
  {2017})\BibitemShut {NoStop}%
\bibitem [{\citenamefont {Mori}\ \emph {et~al.}(2023)\citenamefont {Mori},
  \citenamefont {Yogo}, \citenamefont {Arikawa}, \citenamefont {Hayakawa},
  \citenamefont {Mirfayzi}, \citenamefont {Lan}, \citenamefont {Wei},
  \citenamefont {Abe}, \citenamefont {Nakai} \emph {et~al.}}]{Mori2023}%
  \BibitemOpen
  \bibfield  {author} {\bibinfo {author} {\bibfnamefont {T.}~\bibnamefont
  {Mori}}, \bibinfo {author} {\bibfnamefont {A.}~\bibnamefont {Yogo}}, \bibinfo
  {author} {\bibfnamefont {Y.}~\bibnamefont {Arikawa}}, \bibinfo {author}
  {\bibfnamefont {T.}~\bibnamefont {Hayakawa}}, \bibinfo {author}
  {\bibfnamefont {S.~R.}\ \bibnamefont {Mirfayzi}}, \bibinfo {author}
  {\bibfnamefont {Z.}~\bibnamefont {Lan}}, \bibinfo {author} {\bibfnamefont
  {T.}~\bibnamefont {Wei}}, \bibinfo {author} {\bibfnamefont {Y.}~\bibnamefont
  {Abe}}, \bibinfo {author} {\bibfnamefont {M.}~\bibnamefont {Nakai}}, \emph
  {et~al.},\ }\bibfield  {title} {\bibinfo {title} {Feasibility study of
  laser-driven neutron sources for pharmaceutical applications},\ }\href
  {https://doi.org/10.1017/hpl.2023.4} {\bibfield  {journal} {\bibinfo
  {journal} {High Power Laser Science and Engineering}\ }\textbf {\bibinfo
  {volume} {11}},\ \bibinfo {pages} {e20} (\bibinfo {year} {2023})}\BibitemShut
  {NoStop}%
\bibitem [{\citenamefont {{Zimmer, M.}}\ \emph {et~al.}(2024)\citenamefont
  {{Zimmer, M.}}, \citenamefont {{Rösch, T. F.}}, \citenamefont {{Scheuren,
  S.}}, \citenamefont {{Seupel, T.}}, \citenamefont {{Jäger, T.}},
  \citenamefont {{Kohl, J.}}, \citenamefont {{Hofmann, D.}}, \citenamefont
  {{Schaumann, G.}},\ and\ \citenamefont {{Roth, M.}}}]{Zimmer2024}%
  \BibitemOpen
  \bibfield  {author} {\bibinfo {author} {\bibnamefont {{Zimmer, M.}}},
  \bibinfo {author} {\bibnamefont {{Rösch, T. F.}}}, \bibinfo {author}
  {\bibnamefont {{Scheuren, S.}}}, \bibinfo {author} {\bibnamefont {{Seupel,
  T.}}}, \bibinfo {author} {\bibnamefont {{Jäger, T.}}}, \bibinfo {author}
  {\bibnamefont {{Kohl, J.}}}, \bibinfo {author} {\bibnamefont {{Hofmann,
  D.}}}, \bibinfo {author} {\bibnamefont {{Schaumann, G.}}},\ and\ \bibinfo
  {author} {\bibnamefont {{Roth, M.}}},\ }\bibfield  {title} {\bibinfo {title}
  {Assessing the potential of upcoming laser-driven neutron sources and their
  practical applications for industry and society},\ }\href
  {https://doi.org/10.1140/epjp/s13360-024-05879-5} {\bibfield  {journal}
  {\bibinfo  {journal} {Eur. Phys. J. Plus}\ }\textbf {\bibinfo {volume}
  {139}},\ \bibinfo {pages} {1107} (\bibinfo {year} {2024})}\BibitemShut
  {NoStop}%
\bibitem [{\citenamefont {Mirfayzi}\ \emph {et~al.}(2025)\citenamefont
  {Mirfayzi}, \citenamefont {Gryaznevich}, \citenamefont {Lonsdale},
  \citenamefont {Naylor}, \citenamefont {Takase},\ and\ \citenamefont
  {Kingham}}]{Mirfayzi2025}%
  \BibitemOpen
  \bibfield  {author} {\bibinfo {author} {\bibfnamefont {S.~R.}\ \bibnamefont
  {Mirfayzi}}, \bibinfo {author} {\bibfnamefont {M.}~\bibnamefont
  {Gryaznevich}}, \bibinfo {author} {\bibfnamefont {O.}~\bibnamefont
  {Lonsdale}}, \bibinfo {author} {\bibfnamefont {G.}~\bibnamefont {Naylor}},
  \bibinfo {author} {\bibfnamefont {Y.}~\bibnamefont {Takase}},\ and\ \bibinfo
  {author} {\bibfnamefont {D.}~\bibnamefont {Kingham}},\ }\bibfield  {title}
  {\bibinfo {title} {Recent developments on plasma based neutron sources from
  microscopic innovations to meter-scale applications},\ }\href
  {https://doi.org/10.35848/1347-4065/ad9f00} {\bibfield  {journal} {\bibinfo
  {journal} {Japanese Journal of Applied Physics}\ }\textbf {\bibinfo {volume}
  {64}},\ \bibinfo {pages} {010002} (\bibinfo {year} {2025})}\BibitemShut
  {NoStop}%
\bibitem [{\citenamefont {Canova}\ \emph {et~al.}(2025)\citenamefont {Canova}
  \emph {et~al.}}]{Canova2025}%
  \BibitemOpen
  \bibfield  {author} {\bibinfo {author} {\bibfnamefont {F.}~\bibnamefont
  {Canova}} \emph {et~al.},\ }\bibfield  {title} {\bibinfo {title}
  {Laser-driven fast neutron sources for green energy applications},\
  }\href@noop {} {\bibfield  {journal} {\bibinfo  {journal} {High Power Laser
  Science and Engineeering}\ }\textbf {\bibinfo {volume} {Submitted}},\
  \bibinfo {pages} {0} (\bibinfo {year} {2025})}\BibitemShut {NoStop}%
\bibitem [{\citenamefont {Norreys}\ \emph {et~al.}(1998)\citenamefont
  {Norreys}, \citenamefont {Fews}, \citenamefont {Beg}, \citenamefont {Bell},
  \citenamefont {Dangor}, \citenamefont {Lee}, \citenamefont {Nelson},
  \citenamefont {Schmidt}, \citenamefont {Tatarakis},\ and\ \citenamefont
  {Cable}}]{Norreys1998}%
  \BibitemOpen
  \bibfield  {author} {\bibinfo {author} {\bibfnamefont {P.~A.}\ \bibnamefont
  {Norreys}}, \bibinfo {author} {\bibfnamefont {A.~P.}\ \bibnamefont {Fews}},
  \bibinfo {author} {\bibfnamefont {F.~N.}\ \bibnamefont {Beg}}, \bibinfo
  {author} {\bibfnamefont {A.~R.}\ \bibnamefont {Bell}}, \bibinfo {author}
  {\bibfnamefont {A.~E.}\ \bibnamefont {Dangor}}, \bibinfo {author}
  {\bibfnamefont {P.}~\bibnamefont {Lee}}, \bibinfo {author} {\bibfnamefont
  {M.~B.}\ \bibnamefont {Nelson}}, \bibinfo {author} {\bibfnamefont
  {H.}~\bibnamefont {Schmidt}}, \bibinfo {author} {\bibfnamefont
  {M.}~\bibnamefont {Tatarakis}},\ and\ \bibinfo {author} {\bibfnamefont
  {M.~D.}\ \bibnamefont {Cable}},\ }\bibfield  {title} {\bibinfo {title}
  {Neutron production from picosecond laser irradiation of deuterated targets
  at intensities of \(145^\circ\)},\ }\href
  {https://doi.org/10.1088/0741-3335/40/2/001} {\bibfield  {journal} {\bibinfo
  {journal} {Plasma Physics and Controlled Fusion}\ }\textbf {\bibinfo {volume}
  {40}},\ \bibinfo {pages} {175} (\bibinfo {year} {1998})}\BibitemShut
  {NoStop}%
\bibitem [{\citenamefont {Disdier}\ \emph {et~al.}(1999)\citenamefont
  {Disdier}, \citenamefont {Gar\ifmmode~\mbox{\c{c}}\else \c{c}\fi{}onnet},
  \citenamefont {Malka},\ and\ \citenamefont {Miquel}}]{Disdier1999}%
  \BibitemOpen
  \bibfield  {author} {\bibinfo {author} {\bibfnamefont {L.}~\bibnamefont
  {Disdier}}, \bibinfo {author} {\bibfnamefont {J.-P.}\ \bibnamefont
  {Gar\ifmmode~\mbox{\c{c}}\else \c{c}\fi{}onnet}}, \bibinfo {author}
  {\bibfnamefont {G.}~\bibnamefont {Malka}},\ and\ \bibinfo {author}
  {\bibfnamefont {J.-L.}\ \bibnamefont {Miquel}},\ }\bibfield  {title}
  {\bibinfo {title} {Fast neutron emission from a high-energy ion beam produced
  by a high-intensity subpicosecond laser pulse},\ }\href
  {https://doi.org/10.1103/PhysRevLett.82.1454} {\bibfield  {journal} {\bibinfo
   {journal} {Phys. Rev. Lett.}\ }\textbf {\bibinfo {volume} {82}},\ \bibinfo
  {pages} {1454} (\bibinfo {year} {1999})}\BibitemShut {NoStop}%
\bibitem [{\citenamefont {Ditmire}\ \emph {et~al.}(1999)\citenamefont
  {Ditmire}, \citenamefont {Zweiback}, \citenamefont {Yanovsky}, \citenamefont
  {Cowan}, \citenamefont {Hays},\ and\ \citenamefont {Wharton}}]{Ditmire1999}%
  \BibitemOpen
  \bibfield  {author} {\bibinfo {author} {\bibfnamefont {T.}~\bibnamefont
  {Ditmire}}, \bibinfo {author} {\bibfnamefont {J.}~\bibnamefont {Zweiback}},
  \bibinfo {author} {\bibfnamefont {V.~P.}\ \bibnamefont {Yanovsky}}, \bibinfo
  {author} {\bibfnamefont {T.~E.}\ \bibnamefont {Cowan}}, \bibinfo {author}
  {\bibfnamefont {G.}~\bibnamefont {Hays}},\ and\ \bibinfo {author}
  {\bibfnamefont {K.~B.}\ \bibnamefont {Wharton}},\ }\bibfield  {title}
  {\bibinfo {title} {Nuclear fusion from explosions of femtosecond laser-heated
  deuterium clusters},\ }\href {https://doi.org/10.1038/19037} {\bibfield
  {journal} {\bibinfo  {journal} {Nature}\ }\textbf {\bibinfo {volume} {398}},\
  \bibinfo {pages} {489} (\bibinfo {year} {1999})}\BibitemShut {NoStop}%
\bibitem [{\citenamefont {Kitagawa}\ \emph {et~al.}(2011)\citenamefont
  {Kitagawa}, \citenamefont {Mori}, \citenamefont {Hanayama}, \citenamefont
  {Okihara}, \citenamefont {Fujita}, \citenamefont {Ishii}, \citenamefont
  {Kawashima}, \citenamefont {Sato}, \citenamefont {Sekine}, \citenamefont
  {Yasuhara} \emph {et~al.}}]{Kitagawa2011}%
  \BibitemOpen
  \bibfield  {author} {\bibinfo {author} {\bibfnamefont {Y.}~\bibnamefont
  {Kitagawa}}, \bibinfo {author} {\bibfnamefont {Y.}~\bibnamefont {Mori}},
  \bibinfo {author} {\bibfnamefont {R.}~\bibnamefont {Hanayama}}, \bibinfo
  {author} {\bibfnamefont {S.}~\bibnamefont {Okihara}}, \bibinfo {author}
  {\bibfnamefont {K.}~\bibnamefont {Fujita}}, \bibinfo {author} {\bibfnamefont
  {K.}~\bibnamefont {Ishii}}, \bibinfo {author} {\bibfnamefont
  {T.}~\bibnamefont {Kawashima}}, \bibinfo {author} {\bibfnamefont
  {N.}~\bibnamefont {Sato}}, \bibinfo {author} {\bibfnamefont {T.}~\bibnamefont
  {Sekine}}, \bibinfo {author} {\bibfnamefont {R.}~\bibnamefont {Yasuhara}},
  \emph {et~al.},\ }\bibfield  {title} {\bibinfo {title} {Efficient fusion
  neutron generation using a 10-{TW} high-repetition rate diode-pumped laser},\
  }\href {https://doi.org/10.1585/pfr.6.1306006} {\bibfield  {journal}
  {\bibinfo  {journal} {Plasma and Fusion Research}\ }\textbf {\bibinfo
  {volume} {6}},\ \bibinfo {pages} {1306006} (\bibinfo {year}
  {2011})}\BibitemShut {NoStop}%
\bibitem [{\citenamefont {Bang}\ \emph {et~al.}(2013)\citenamefont {Bang},
  \citenamefont {Dyer}, \citenamefont {Quevedo}, \citenamefont {Bernstein},
  \citenamefont {Gaul}, \citenamefont {Donovan},\ and\ \citenamefont
  {Ditmire}}]{Wang2013}%
  \BibitemOpen
  \bibfield  {author} {\bibinfo {author} {\bibfnamefont {W.}~\bibnamefont
  {Bang}}, \bibinfo {author} {\bibfnamefont {G.}~\bibnamefont {Dyer}}, \bibinfo
  {author} {\bibfnamefont {H.~J.}\ \bibnamefont {Quevedo}}, \bibinfo {author}
  {\bibfnamefont {A.~C.}\ \bibnamefont {Bernstein}}, \bibinfo {author}
  {\bibfnamefont {E.}~\bibnamefont {Gaul}}, \bibinfo {author} {\bibfnamefont
  {M.}~\bibnamefont {Donovan}},\ and\ \bibinfo {author} {\bibfnamefont
  {T.}~\bibnamefont {Ditmire}},\ }\bibfield  {title} {\bibinfo {title}
  {Optimization of the neutron yield in fusion plasmas produced by {C}oulomb
  explosions of deuterium clusters irradiated by a petawatt laser},\ }\href
  {https://doi.org/10.1103/PhysRevE.87.023106} {\bibfield  {journal} {\bibinfo
  {journal} {Phys. Rev. E}\ }\textbf {\bibinfo {volume} {87}},\ \bibinfo
  {pages} {023106} (\bibinfo {year} {2013})}\BibitemShut {NoStop}%
\bibitem [{\citenamefont {Willingale}\ \emph {et~al.}(2011)\citenamefont
  {Willingale}, \citenamefont {Petrov}, \citenamefont {Maksimchuk},
  \citenamefont {Davis}, \citenamefont {Freeman}, \citenamefont {Joglekar},
  \citenamefont {Matsuoka}, \citenamefont {Murphy}, \citenamefont
  {Ovchinnikov}, \citenamefont {Thomas} \emph {et~al.}}]{Willingale2011}%
  \BibitemOpen
  \bibfield  {author} {\bibinfo {author} {\bibfnamefont {L.}~\bibnamefont
  {Willingale}}, \bibinfo {author} {\bibfnamefont {G.~M.}\ \bibnamefont
  {Petrov}}, \bibinfo {author} {\bibfnamefont {A.}~\bibnamefont {Maksimchuk}},
  \bibinfo {author} {\bibfnamefont {J.}~\bibnamefont {Davis}}, \bibinfo
  {author} {\bibfnamefont {R.~R.}\ \bibnamefont {Freeman}}, \bibinfo {author}
  {\bibfnamefont {A.~S.}\ \bibnamefont {Joglekar}}, \bibinfo {author}
  {\bibfnamefont {T.}~\bibnamefont {Matsuoka}}, \bibinfo {author}
  {\bibfnamefont {C.~D.}\ \bibnamefont {Murphy}}, \bibinfo {author}
  {\bibfnamefont {V.~M.}\ \bibnamefont {Ovchinnikov}}, \bibinfo {author}
  {\bibfnamefont {A.~G.~R.}\ \bibnamefont {Thomas}}, \emph {et~al.},\
  }\bibfield  {title} {\bibinfo {title} {Comparison of bulk and pitcher-catcher
  targets for laser-driven neutron production},\ }\bibfield  {journal}
  {\bibinfo  {journal} {Physics of Plasmas}\ }\textbf {\bibinfo {volume}
  {18}},\ \href {https://doi.org/10.1063/1.3624769} {10.1063/1.3624769}
  (\bibinfo {year} {2011})\BibitemShut {NoStop}%
\bibitem [{\citenamefont {Günther}\ \emph {et~al.}(2022)\citenamefont
  {Günther}, \citenamefont {Rosmej}, \citenamefont {Tavana}, \citenamefont
  {Gyrdymov}, \citenamefont {Skobliakov}, \citenamefont {Kantsyrev},
  \citenamefont {Zähter}, \citenamefont {Borisenko}, \citenamefont {Pukhov},\
  and\ \citenamefont {Andreev}}]{Guenther2022}%
  \BibitemOpen
  \bibfield  {author} {\bibinfo {author} {\bibfnamefont {M.~M.}\ \bibnamefont
  {Günther}}, \bibinfo {author} {\bibfnamefont {O.~N.}\ \bibnamefont
  {Rosmej}}, \bibinfo {author} {\bibfnamefont {P.}~\bibnamefont {Tavana}},
  \bibinfo {author} {\bibfnamefont {M.}~\bibnamefont {Gyrdymov}}, \bibinfo
  {author} {\bibfnamefont {A.}~\bibnamefont {Skobliakov}}, \bibinfo {author}
  {\bibfnamefont {A.}~\bibnamefont {Kantsyrev}}, \bibinfo {author}
  {\bibfnamefont {S.}~\bibnamefont {Zähter}}, \bibinfo {author} {\bibfnamefont
  {N.~G.}\ \bibnamefont {Borisenko}}, \bibinfo {author} {\bibfnamefont
  {A.}~\bibnamefont {Pukhov}},\ and\ \bibinfo {author} {\bibfnamefont {N.~E.}\
  \bibnamefont {Andreev}},\ }\bibfield  {title} {\bibinfo {title}
  {Forward-looking insights in laser-generated ultra-intense $\gamma$-ray and
  neutron sources for nuclear application and science},\ }\href
  {https://doi.org/10.1038/s41467-021-27694-7} {\bibfield  {journal} {\bibinfo
  {journal} {Nature Communications}\ }\textbf {\bibinfo {volume} {13}},\
  \bibinfo {pages} {170} (\bibinfo {year} {2022})}\BibitemShut {NoStop}%
\bibitem [{\citenamefont {Yogo}\ \emph
  {et~al.}(2023{\natexlab{b}})\citenamefont {Yogo}, \citenamefont {Lan},
  \citenamefont {Arikawa}, \citenamefont {Abe}, \citenamefont {Mirfayzi},
  \citenamefont {Wei}, \citenamefont {Mori}, \citenamefont {Golovin},
  \citenamefont {Hayakawa}, \citenamefont {Iwata} \emph {et~al.}}]{Yogo2023}%
  \BibitemOpen
  \bibfield  {author} {\bibinfo {author} {\bibfnamefont {A.}~\bibnamefont
  {Yogo}}, \bibinfo {author} {\bibfnamefont {Z.}~\bibnamefont {Lan}}, \bibinfo
  {author} {\bibfnamefont {Y.}~\bibnamefont {Arikawa}}, \bibinfo {author}
  {\bibfnamefont {Y.}~\bibnamefont {Abe}}, \bibinfo {author} {\bibfnamefont
  {S.~R.}\ \bibnamefont {Mirfayzi}}, \bibinfo {author} {\bibfnamefont
  {T.}~\bibnamefont {Wei}}, \bibinfo {author} {\bibfnamefont {T.}~\bibnamefont
  {Mori}}, \bibinfo {author} {\bibfnamefont {D.}~\bibnamefont {Golovin}},
  \bibinfo {author} {\bibfnamefont {T.}~\bibnamefont {Hayakawa}}, \bibinfo
  {author} {\bibfnamefont {N.}~\bibnamefont {Iwata}}, \emph {et~al.},\
  }\bibfield  {title} {\bibinfo {title} {Laser-driven neutron generation
  realizing single-shot resonance spectroscopy},\ }\href
  {https://doi.org/10.1103/PhysRevX.13.011011} {\bibfield  {journal} {\bibinfo
  {journal} {Phys. Rev. X}\ }\textbf {\bibinfo {volume} {13}},\ \bibinfo
  {pages} {011011} (\bibinfo {year} {2023}{\natexlab{b}})}\BibitemShut
  {NoStop}%
\bibitem [{\citenamefont {Prencipe}\ \emph {et~al.}(2017)\citenamefont
  {Prencipe}, \citenamefont {Fuchs}, \citenamefont {Pascarelli} \emph
  {et~al.}}]{Prencipe2017}%
  \BibitemOpen
  \bibfield  {author} {\bibinfo {author} {\bibfnamefont {I.}~\bibnamefont
  {Prencipe}}, \bibinfo {author} {\bibfnamefont {J.}~\bibnamefont {Fuchs}},
  \bibinfo {author} {\bibfnamefont {S.}~\bibnamefont {Pascarelli}}, \emph
  {et~al.},\ }\bibfield  {title} {\bibinfo {title} {Targets for high repetition
  rate laser facilities: needs, challenges and perspectives},\ }\href
  {https://doi.org/10.1017/hpl.2017.18} {\bibfield  {journal} {\bibinfo
  {journal} {High Power Laser Science and Engineering}\ }\textbf {\bibinfo
  {volume} {5}},\ \bibinfo {pages} {e17} (\bibinfo {year} {2017})}\BibitemShut
  {NoStop}%
\bibitem [{\citenamefont {Osvay}\ \emph
  {et~al.}(2024{\natexlab{a}})\citenamefont {Osvay}, \citenamefont {Singh},
  \citenamefont {Varmazyar}, \citenamefont {F{\"u}le}, \citenamefont
  {Gilinger}, \citenamefont {Kis}, \citenamefont {Lehotai}, \citenamefont
  {Nagy}, \citenamefont {Stuhl}, \citenamefont {Elekes} \emph
  {et~al.}}]{Osvay2024_1Hz}%
  \BibitemOpen
  \bibfield  {author} {\bibinfo {author} {\bibfnamefont {K.}~\bibnamefont
  {Osvay}}, \bibinfo {author} {\bibfnamefont {P.~K.}\ \bibnamefont {Singh}},
  \bibinfo {author} {\bibfnamefont {P.}~\bibnamefont {Varmazyar}}, \bibinfo
  {author} {\bibfnamefont {M.}~\bibnamefont {F{\"u}le}}, \bibinfo {author}
  {\bibfnamefont {T.}~\bibnamefont {Gilinger}}, \bibinfo {author}
  {\bibfnamefont {B.}~\bibnamefont {Kis}}, \bibinfo {author} {\bibfnamefont
  {L.}~\bibnamefont {Lehotai}}, \bibinfo {author} {\bibfnamefont
  {B.}~\bibnamefont {Nagy}}, \bibinfo {author} {\bibfnamefont {L.}~\bibnamefont
  {Stuhl}}, \bibinfo {author} {\bibfnamefont {Z.}~\bibnamefont {Elekes}}, \emph
  {et~al.},\ }\bibfield  {title} {\bibinfo {title} {Fast neutron generation
  with few-cycle, relativistic laser pulses at 1{H}z repetition rate},\ }\href
  {https://doi.org/10.1038/s41598-024-75855-7} {\bibfield  {journal} {\bibinfo
  {journal} {Scientific Reports}\ }\textbf {\bibinfo {volume} {14}},\ \bibinfo
  {pages} {25302} (\bibinfo {year} {2024}{\natexlab{a}})}\BibitemShut {NoStop}%
\bibitem [{\citenamefont {Lelièvre}\ \emph {et~al.}(2024)\citenamefont
  {Lelièvre}, \citenamefont {Catrix}, \citenamefont {Vallières},
  \citenamefont {Fourmaux}, \citenamefont {Allaoua}, \citenamefont
  {Anthonippillai}, \citenamefont {Antici}, \citenamefont {Ducasse},\ and\
  \citenamefont {Fuchs}}]{Lelievre2024}%
  \BibitemOpen
  \bibfield  {author} {\bibinfo {author} {\bibfnamefont {R.}~\bibnamefont
  {Lelièvre}}, \bibinfo {author} {\bibfnamefont {E.}~\bibnamefont {Catrix}},
  \bibinfo {author} {\bibfnamefont {S.}~\bibnamefont {Vallières}}, \bibinfo
  {author} {\bibfnamefont {S.}~\bibnamefont {Fourmaux}}, \bibinfo {author}
  {\bibfnamefont {A.}~\bibnamefont {Allaoua}}, \bibinfo {author} {\bibfnamefont
  {V.}~\bibnamefont {Anthonippillai}}, \bibinfo {author} {\bibfnamefont
  {P.}~\bibnamefont {Antici}}, \bibinfo {author} {\bibfnamefont
  {Q.}~\bibnamefont {Ducasse}},\ and\ \bibinfo {author} {\bibfnamefont
  {J.}~\bibnamefont {Fuchs}},\ }\bibfield  {title} {\bibinfo {title} {High
  repetition-rate 0.5 {H}z broadband neutron source driven by the advanced
  laser light source},\ }\href {https://doi.org/10.1063/5.0218582} {\bibfield
  {journal} {\bibinfo  {journal} {Physics of Plasmas}\ }\textbf {\bibinfo
  {volume} {31}},\ \bibinfo {pages} {093106} (\bibinfo {year}
  {2024})}\BibitemShut {NoStop}%
\bibitem [{\citenamefont {{Osvay, K.}}\ \emph {et~al.}(2024)\citenamefont
  {{Osvay, K.}}, \citenamefont {{Stuhl, L.}}, \citenamefont {{Varmazyar, P.}},
  \citenamefont {{Gilinger, T.}}, \citenamefont {{Elekes, Z.}}, \citenamefont
  {{Fenyvesi, A.}}, \citenamefont {{Hideghethy, K.}}, \citenamefont {{Szabo, R.
  E.}}, \citenamefont {{F{\"u}le, M.}}, \citenamefont {{Biró, B.}} \emph
  {et~al.}}]{Osvay2024towards}%
  \BibitemOpen
  \bibfield  {author} {\bibinfo {author} {\bibnamefont {{Osvay, K.}}}, \bibinfo
  {author} {\bibnamefont {{Stuhl, L.}}}, \bibinfo {author} {\bibnamefont
  {{Varmazyar, P.}}}, \bibinfo {author} {\bibnamefont {{Gilinger, T.}}},
  \bibinfo {author} {\bibnamefont {{Elekes, Z.}}}, \bibinfo {author}
  {\bibnamefont {{Fenyvesi, A.}}}, \bibinfo {author} {\bibnamefont
  {{Hideghethy, K.}}}, \bibinfo {author} {\bibnamefont {{Szabo, R. E.}}},
  \bibinfo {author} {\bibnamefont {{F{\"u}le, M.}}}, \bibinfo {author}
  {\bibnamefont {{Biró, B.}}}, \emph {et~al.},\ }\bibfield  {title} {\bibinfo
  {title} {Towards a 10\textsuperscript{10} n/s neutron source with k{H}z
  repetition rate, few-cycle laser pulses},\ }\href
  {https://doi.org/10.1140/epjp/s13360-024-05338-1} {\bibfield  {journal}
  {\bibinfo  {journal} {Eur. Phys. J. Plus}\ }\textbf {\bibinfo {volume}
  {139}},\ \bibinfo {pages} {574} (\bibinfo {year} {2024})}\BibitemShut
  {NoStop}%
\bibitem [{\citenamefont {Treffert}\ \emph {et~al.}(2021)\citenamefont
  {Treffert}, \citenamefont {Curry}, \citenamefont {Ditmire}, \citenamefont
  {Glenn}, \citenamefont {Quevedo}, \citenamefont {Roth}, \citenamefont
  {Schoenwaelder}, \citenamefont {Zimmer}, \citenamefont {Glenzer},\ and\
  \citenamefont {Gauthier}}]{treffert2021towards}%
  \BibitemOpen
  \bibfield  {author} {\bibinfo {author} {\bibfnamefont {F.}~\bibnamefont
  {Treffert}}, \bibinfo {author} {\bibfnamefont {C.~B.}\ \bibnamefont {Curry}},
  \bibinfo {author} {\bibfnamefont {T.}~\bibnamefont {Ditmire}}, \bibinfo
  {author} {\bibfnamefont {G.~D.}\ \bibnamefont {Glenn}}, \bibinfo {author}
  {\bibfnamefont {H.~J.}\ \bibnamefont {Quevedo}}, \bibinfo {author}
  {\bibfnamefont {M.}~\bibnamefont {Roth}}, \bibinfo {author} {\bibfnamefont
  {C.}~\bibnamefont {Schoenwaelder}}, \bibinfo {author} {\bibfnamefont
  {M.}~\bibnamefont {Zimmer}}, \bibinfo {author} {\bibfnamefont {S.~H.}\
  \bibnamefont {Glenzer}},\ and\ \bibinfo {author} {\bibfnamefont
  {M.}~\bibnamefont {Gauthier}},\ }\bibfield  {title} {\bibinfo {title}
  {Towards high-repetition-rate fast neutron sources using novel enabling
  technologies},\ }\href {https://doi.org/10.3390/instruments5040038}
  {\bibfield  {journal} {\bibinfo  {journal} {Instruments}\ }\textbf {\bibinfo
  {volume} {5}},\ \bibinfo {pages} {38} (\bibinfo {year} {2021})}\BibitemShut
  {NoStop}%
\bibitem [{\citenamefont {Hah}\ \emph {et~al.}(2016)\citenamefont {Hah},
  \citenamefont {Petrov}, \citenamefont {Nees}, \citenamefont {He},
  \citenamefont {Hammig}, \citenamefont {Krushelnick},\ and\ \citenamefont
  {Thomas}}]{Hah2016}%
  \BibitemOpen
  \bibfield  {author} {\bibinfo {author} {\bibfnamefont {J.}~\bibnamefont
  {Hah}}, \bibinfo {author} {\bibfnamefont {G.~M.}\ \bibnamefont {Petrov}},
  \bibinfo {author} {\bibfnamefont {J.~A.}\ \bibnamefont {Nees}}, \bibinfo
  {author} {\bibfnamefont {Z.-H.}\ \bibnamefont {He}}, \bibinfo {author}
  {\bibfnamefont {M.~D.}\ \bibnamefont {Hammig}}, \bibinfo {author}
  {\bibfnamefont {K.}~\bibnamefont {Krushelnick}},\ and\ \bibinfo {author}
  {\bibfnamefont {A.~G.~R.}\ \bibnamefont {Thomas}},\ }\bibfield  {title}
  {\bibinfo {title} {High repetition-rate neutron generation by several-{mJ},
  35 fs pulses interacting with free-flowing {D}$_2${O}},\ }\href
  {https://doi.org/10.1063/1.4963819} {\bibfield  {journal} {\bibinfo
  {journal} {Applied Physics Letters}\ }\textbf {\bibinfo {volume} {109}},\
  \bibinfo {pages} {144102} (\bibinfo {year} {2016})}\BibitemShut {NoStop}%
\bibitem [{\citenamefont {Hah}\ \emph {et~al.}(2018)\citenamefont {Hah},
  \citenamefont {Nees}, \citenamefont {Hammig}, \citenamefont {Krushelnick},\
  and\ \citenamefont {Thomas}}]{Hah2018}%
  \BibitemOpen
  \bibfield  {author} {\bibinfo {author} {\bibfnamefont {J.}~\bibnamefont
  {Hah}}, \bibinfo {author} {\bibfnamefont {J.~A.}\ \bibnamefont {Nees}},
  \bibinfo {author} {\bibfnamefont {M.~D.}\ \bibnamefont {Hammig}}, \bibinfo
  {author} {\bibfnamefont {K.}~\bibnamefont {Krushelnick}},\ and\ \bibinfo
  {author} {\bibfnamefont {A.~G.~R.}\ \bibnamefont {Thomas}},\ }\bibfield
  {title} {\bibinfo {title} {Characterization of a high repetition-rate
  laser-driven short-pulsed neutron source},\ }\href
  {https://doi.org/10.1088/1361-6587/aab327} {\bibfield  {journal} {\bibinfo
  {journal} {Plasma Physics and Controlled Fusion}\ }\textbf {\bibinfo {volume}
  {60}},\ \bibinfo {pages} {054011} (\bibinfo {year} {2018})}\BibitemShut
  {NoStop}%
\bibitem [{\citenamefont {Knight}\ \emph {et~al.}(2024)\citenamefont {Knight},
  \citenamefont {Gautam}, \citenamefont {Stoner} \emph {et~al.}}]{Knight2024}%
  \BibitemOpen
  \bibfield  {author} {\bibinfo {author} {\bibfnamefont {B.~M.}\ \bibnamefont
  {Knight}}, \bibinfo {author} {\bibfnamefont {C.~M.}\ \bibnamefont {Gautam}},
  \bibinfo {author} {\bibfnamefont {C.~R.}\ \bibnamefont {Stoner}}, \emph
  {et~al.},\ }\bibfield  {title} {\bibinfo {title} {Detailed characterization
  of k{H}z-rate laser-driven fusion at a thin liquid sheet with a neutron
  detection suite},\ }\href {https://doi.org/10.1017/hpl.2023.84} {\bibfield
  {journal} {\bibinfo  {journal} {High Power Laser Science and Engineering}\
  }\textbf {\bibinfo {volume} {12}},\ \bibinfo {pages} {e2} (\bibinfo {year}
  {2024})}\BibitemShut {NoStop}%
\bibitem [{\citenamefont {Treffert}\ \emph {et~al.}(2022)\citenamefont
  {Treffert}, \citenamefont {Curry}, \citenamefont {Chou}, \citenamefont
  {Crissman}, \citenamefont {DePonte}, \citenamefont {Fiuza}, \citenamefont
  {Glenn}, \citenamefont {Hollinger}, \citenamefont {Nedbailo}, \citenamefont
  {Park} \emph {et~al.}}]{Treffert2022}%
  \BibitemOpen
  \bibfield  {author} {\bibinfo {author} {\bibfnamefont {F.}~\bibnamefont
  {Treffert}}, \bibinfo {author} {\bibfnamefont {C.~B.}\ \bibnamefont {Curry}},
  \bibinfo {author} {\bibfnamefont {H.-G.~J.}\ \bibnamefont {Chou}}, \bibinfo
  {author} {\bibfnamefont {C.~J.}\ \bibnamefont {Crissman}}, \bibinfo {author}
  {\bibfnamefont {D.~P.}\ \bibnamefont {DePonte}}, \bibinfo {author}
  {\bibfnamefont {F.}~\bibnamefont {Fiuza}}, \bibinfo {author} {\bibfnamefont
  {G.~D.}\ \bibnamefont {Glenn}}, \bibinfo {author} {\bibfnamefont {R.~C.}\
  \bibnamefont {Hollinger}}, \bibinfo {author} {\bibfnamefont {R.}~\bibnamefont
  {Nedbailo}}, \bibinfo {author} {\bibfnamefont {J.}~\bibnamefont {Park}},
  \emph {et~al.},\ }\bibfield  {title} {\bibinfo {title} {High-repetition-rate,
  multi-{M}e{V} deuteron acceleration from converging heavy water microjets at
  laser intensities of 10$^{21}$ {W}/cm$^2$},\ }\href
  {https://doi.org/10.1063/5.0098973} {\bibfield  {journal} {\bibinfo
  {journal} {Applied Physics Letters}\ }\textbf {\bibinfo {volume} {121}},\
  \bibinfo {pages} {074104} (\bibinfo {year} {2022})}\BibitemShut {NoStop}%
\bibitem [{\citenamefont {F{\"u}le}\ \emph {et~al.}(2024)\citenamefont
  {F{\"u}le}, \citenamefont {Kov{\'a}cs}, \citenamefont {Gilinger},
  \citenamefont {Karnok}, \citenamefont {Ga{\'a}l}, \citenamefont {Figul},
  \citenamefont {Marowsky},\ and\ \citenamefont {Osvay}}]{Fule2024}%
  \BibitemOpen
  \bibfield  {author} {\bibinfo {author} {\bibfnamefont {M.}~\bibnamefont
  {F{\"u}le}}, \bibinfo {author} {\bibfnamefont {A.~P.}\ \bibnamefont
  {Kov{\'a}cs}}, \bibinfo {author} {\bibfnamefont {T.}~\bibnamefont
  {Gilinger}}, \bibinfo {author} {\bibfnamefont {M.}~\bibnamefont {Karnok}},
  \bibinfo {author} {\bibfnamefont {P.}~\bibnamefont {Ga{\'a}l}}, \bibinfo
  {author} {\bibfnamefont {S.}~\bibnamefont {Figul}}, \bibinfo {author}
  {\bibfnamefont {G.}~\bibnamefont {Marowsky}},\ and\ \bibinfo {author}
  {\bibfnamefont {K.}~\bibnamefont {Osvay}},\ }\bibfield  {title} {\bibinfo
  {title} {Development of an ultrathin liquid sheet target for laser ion
  acceleration at high repetition rates in the kilohertz range},\ }\href
  {https://doi.org/10.1017/hpl.2024.19} {\bibfield  {journal} {\bibinfo
  {journal} {High Power Laser Science and Engineering}\ }\textbf {\bibinfo
  {volume} {12}},\ \bibinfo {pages} {e37} (\bibinfo {year} {2024})}\BibitemShut
  {NoStop}%
\bibitem [{\citenamefont {Toth}\ \emph {et~al.}(2020)\citenamefont {Toth},
  \citenamefont {Stanislauskas}, \citenamefont {Balciunas}, \citenamefont
  {Budriunas}, \citenamefont {Adamonis}, \citenamefont {Danilevicius},
  \citenamefont {Viskontas}, \citenamefont {Lengvinas}, \citenamefont {Veitas},
  \citenamefont {Gadonas} \emph {et~al.}}]{Toth2020}%
  \BibitemOpen
  \bibfield  {author} {\bibinfo {author} {\bibfnamefont {S.}~\bibnamefont
  {Toth}}, \bibinfo {author} {\bibfnamefont {T.}~\bibnamefont {Stanislauskas}},
  \bibinfo {author} {\bibfnamefont {I.}~\bibnamefont {Balciunas}}, \bibinfo
  {author} {\bibfnamefont {R.}~\bibnamefont {Budriunas}}, \bibinfo {author}
  {\bibfnamefont {J.}~\bibnamefont {Adamonis}}, \bibinfo {author}
  {\bibfnamefont {R.}~\bibnamefont {Danilevicius}}, \bibinfo {author}
  {\bibfnamefont {K.}~\bibnamefont {Viskontas}}, \bibinfo {author}
  {\bibfnamefont {D.}~\bibnamefont {Lengvinas}}, \bibinfo {author}
  {\bibfnamefont {G.}~\bibnamefont {Veitas}}, \bibinfo {author} {\bibfnamefont
  {D.}~\bibnamefont {Gadonas}}, \emph {et~al.},\ }\bibfield  {title} {\bibinfo
  {title} {Sylos lasers – the frontier of few-cycle, multi-tw, khz lasers for
  ultrafast applications at extreme light infrastructure attosecond light pulse
  source},\ }\href {https://doi.org/10.1088/2515-7647/ab9fe1} {\bibfield
  {journal} {\bibinfo  {journal} {Journal of Physics: Photonics}\ }\textbf
  {\bibinfo {volume} {2}},\ \bibinfo {pages} {045003} (\bibinfo {year}
  {2020})}\BibitemShut {NoStop}%
\bibitem [{\citenamefont {Stuhl}\ \emph {et~al.}(2020)\citenamefont {Stuhl},
  \citenamefont {Sasano}, \citenamefont {Gao}, \citenamefont {Hirai},
  \citenamefont {Yako}, \citenamefont {Wakasa}, \citenamefont {Ahn},
  \citenamefont {Baba}, \citenamefont {Chilug}, \citenamefont {Franchoo} \emph
  {et~al.}}]{Stuhl2020}%
  \BibitemOpen
  \bibfield  {author} {\bibinfo {author} {\bibfnamefont {L.}~\bibnamefont
  {Stuhl}}, \bibinfo {author} {\bibfnamefont {M.}~\bibnamefont {Sasano}},
  \bibinfo {author} {\bibfnamefont {J.}~\bibnamefont {Gao}}, \bibinfo {author}
  {\bibfnamefont {Y.}~\bibnamefont {Hirai}}, \bibinfo {author} {\bibfnamefont
  {K.}~\bibnamefont {Yako}}, \bibinfo {author} {\bibfnamefont {T.}~\bibnamefont
  {Wakasa}}, \bibinfo {author} {\bibfnamefont {D.}~\bibnamefont {Ahn}},
  \bibinfo {author} {\bibfnamefont {H.}~\bibnamefont {Baba}}, \bibinfo {author}
  {\bibfnamefont {A.}~\bibnamefont {Chilug}}, \bibinfo {author} {\bibfnamefont
  {S.}~\bibnamefont {Franchoo}}, \emph {et~al.},\ }\bibfield  {title} {\bibinfo
  {title} {Study of spin-isospin responses of radioactive nuclei with the
  background-reduced neutron spectrometer, pandora},\ }\href
  {https://doi.org/https://doi.org/10.1016/j.nimb.2019.05.057} {\bibfield
  {journal} {\bibinfo  {journal} {Nuclear Instruments and Methods B}\ }\textbf
  {\bibinfo {volume} {463}},\ \bibinfo {pages} {189} (\bibinfo {year}
  {2020})}\BibitemShut {NoStop}%
\bibitem [{\citenamefont {Ziegler}\ \emph {et~al.}(2021)\citenamefont
  {Ziegler}, \citenamefont {Albach}, \citenamefont {Bernert}, \citenamefont
  {Bock}, \citenamefont {Brack}, \citenamefont {Cowan}, \citenamefont {Dover},
  \citenamefont {Garten}, \citenamefont {Gaus}, \citenamefont {Gebhardt} \emph
  {et~al.}}]{Ziegler2021}%
  \BibitemOpen
  \bibfield  {author} {\bibinfo {author} {\bibfnamefont {T.}~\bibnamefont
  {Ziegler}}, \bibinfo {author} {\bibfnamefont {D.}~\bibnamefont {Albach}},
  \bibinfo {author} {\bibfnamefont {C.}~\bibnamefont {Bernert}}, \bibinfo
  {author} {\bibfnamefont {S.}~\bibnamefont {Bock}}, \bibinfo {author}
  {\bibfnamefont {F.-E.}\ \bibnamefont {Brack}}, \bibinfo {author}
  {\bibfnamefont {T.~E.}\ \bibnamefont {Cowan}}, \bibinfo {author}
  {\bibfnamefont {N.~P.}\ \bibnamefont {Dover}}, \bibinfo {author}
  {\bibfnamefont {M.}~\bibnamefont {Garten}}, \bibinfo {author} {\bibfnamefont
  {L.}~\bibnamefont {Gaus}}, \bibinfo {author} {\bibfnamefont {R.}~\bibnamefont
  {Gebhardt}}, \emph {et~al.},\ }\bibfield  {title} {\bibinfo {title} {Proton
  beam quality enhancement by spectral phase control of a {PW}-class laser
  system},\ }\href {https://doi.org/10.1038/s41598-021-86547-x} {\bibfield
  {journal} {\bibinfo  {journal} {Scientific Reports}\ }\textbf {\bibinfo
  {volume} {11}},\ \bibinfo {pages} {7338} (\bibinfo {year}
  {2021})}\BibitemShut {NoStop}%
\bibitem [{\citenamefont {Osvay}\ \emph
  {et~al.}(2024{\natexlab{b}})\citenamefont {Osvay} \emph
  {et~al.}}]{Osvay2024AAC}%
  \BibitemOpen
  \bibfield  {author} {\bibinfo {author} {\bibfnamefont {K.}~\bibnamefont
  {Osvay}} \emph {et~al.},\ }\bibfield  {title} {\bibinfo {title} {Enhancing of
  deuteron acceleration with spectral phase modulation of few-cycle laser
  pulses},\ }in\ \href@noop {} {\emph {\bibinfo {booktitle} {Advanced
  Accelerator Concept Workshop, Naperville, IL, USA}}}\ (\bibinfo {year}
  {2024})\BibitemShut {NoStop}%
\bibitem [{\citenamefont {Schreiber}\ \emph {et~al.}(2014)\citenamefont
  {Schreiber}, \citenamefont {Bell},\ and\ \citenamefont
  {Najmudin}}]{Schreiber2014}%
  \BibitemOpen
  \bibfield  {author} {\bibinfo {author} {\bibfnamefont {J.}~\bibnamefont
  {Schreiber}}, \bibinfo {author} {\bibfnamefont {F.}~\bibnamefont {Bell}},\
  and\ \bibinfo {author} {\bibfnamefont {Z.}~\bibnamefont {Najmudin}},\
  }\bibfield  {title} {\bibinfo {title} {Optimization of relativistic
  laser–ion acceleration},\ }\href {https://doi.org/doi:10.1017/hpl.2014.46}
  {\bibfield  {journal} {\bibinfo  {journal} {High Power Laser Science and
  Engineering}\ }\textbf {\bibinfo {volume} {2}},\ \bibinfo {pages} {e41}
  (\bibinfo {year} {2014})}\BibitemShut {NoStop}%
\bibitem [{\citenamefont {Agostinelli}\ \emph {et~al.}(2003)\citenamefont
  {Agostinelli}, \citenamefont {Allison}, \citenamefont {Amako}, \citenamefont
  {Apostolakis}, \citenamefont {Araujo}, \citenamefont {Arce}, \citenamefont
  {Asai}, \citenamefont {Axen}, \citenamefont {Banerjee}, \citenamefont
  {Barrand} \emph {et~al.}}]{geant4}%
  \BibitemOpen
  \bibfield  {author} {\bibinfo {author} {\bibfnamefont {S.}~\bibnamefont
  {Agostinelli}}, \bibinfo {author} {\bibfnamefont {J.}~\bibnamefont
  {Allison}}, \bibinfo {author} {\bibfnamefont {K.}~\bibnamefont {Amako}},
  \bibinfo {author} {\bibfnamefont {J.}~\bibnamefont {Apostolakis}}, \bibinfo
  {author} {\bibfnamefont {H.}~\bibnamefont {Araujo}}, \bibinfo {author}
  {\bibfnamefont {P.}~\bibnamefont {Arce}}, \bibinfo {author} {\bibfnamefont
  {M.}~\bibnamefont {Asai}}, \bibinfo {author} {\bibfnamefont {D.}~\bibnamefont
  {Axen}}, \bibinfo {author} {\bibfnamefont {S.}~\bibnamefont {Banerjee}},
  \bibinfo {author} {\bibfnamefont {G.}~\bibnamefont {Barrand}}, \emph
  {et~al.},\ }\bibfield  {title} {\bibinfo {title} {Geant4—a simulation
  toolkit},\ }\href
  {https://doi.org/https://doi.org/10.1016/S0168-9002(03)01368-8} {\bibfield
  {journal} {\bibinfo  {journal} {Nuclear Instruments and Methods A}\ }\textbf
  {\bibinfo {volume} {506}},\ \bibinfo {pages} {250} (\bibinfo {year}
  {2003})}\BibitemShut {NoStop}%
\bibitem [{\citenamefont {Cao}\ \emph {et~al.}(2023)\citenamefont {Cao},
  \citenamefont {Peng}, \citenamefont {Shou}, \citenamefont {Zhao},
  \citenamefont {Chen}, \citenamefont {Gao}, \citenamefont {Liu}, \citenamefont
  {Wang}, \citenamefont {Mei}, \citenamefont {Pan} \emph
  {et~al.}}]{CaoFrontiers2023}%
  \BibitemOpen
  \bibfield  {author} {\bibinfo {author} {\bibfnamefont {Z.}~\bibnamefont
  {Cao}}, \bibinfo {author} {\bibfnamefont {Z.}~\bibnamefont {Peng}}, \bibinfo
  {author} {\bibfnamefont {Y.}~\bibnamefont {Shou}}, \bibinfo {author}
  {\bibfnamefont {J.}~\bibnamefont {Zhao}}, \bibinfo {author} {\bibfnamefont
  {S.}~\bibnamefont {Chen}}, \bibinfo {author} {\bibfnamefont {Y.}~\bibnamefont
  {Gao}}, \bibinfo {author} {\bibfnamefont {J.}~\bibnamefont {Liu}}, \bibinfo
  {author} {\bibfnamefont {P.}~\bibnamefont {Wang}}, \bibinfo {author}
  {\bibfnamefont {Z.}~\bibnamefont {Mei}}, \bibinfo {author} {\bibfnamefont
  {Z.}~\bibnamefont {Pan}}, \emph {et~al.},\ }\bibfield  {title} {\bibinfo
  {title} {Vibration and jitter of free-flowing thin liquid sheets as target
  for high-repetition-rate laser-ion acceleration},\ }\href
  {https://doi.org/10.3389/fphy.2023.1172075} {\bibfield  {journal} {\bibinfo
  {journal} {Frontiers in Physics}\ }\textbf {\bibinfo {volume} {11}},\
  \bibinfo {pages} {1172075} (\bibinfo {year} {2023})}\BibitemShut {NoStop}%
\bibitem [{\citenamefont {Peng}\ \emph {et~al.}(2024)\citenamefont {Peng},
  \citenamefont {Cao}, \citenamefont {Liu}, \citenamefont {Shou}, \citenamefont
  {Zhao}, \citenamefont {Chen}, \citenamefont {Gao}, \citenamefont {Wang},
  \citenamefont {Mei}, \citenamefont {Pan},\ and\ \citenamefont
  {et~al.}}]{Peng_HPLSE_2024}%
  \BibitemOpen
  \bibfield  {author} {\bibinfo {author} {\bibfnamefont {Z.}~\bibnamefont
  {Peng}}, \bibinfo {author} {\bibfnamefont {Z.}~\bibnamefont {Cao}}, \bibinfo
  {author} {\bibfnamefont {X.}~\bibnamefont {Liu}}, \bibinfo {author}
  {\bibfnamefont {Y.}~\bibnamefont {Shou}}, \bibinfo {author} {\bibfnamefont
  {J.}~\bibnamefont {Zhao}}, \bibinfo {author} {\bibfnamefont {S.}~\bibnamefont
  {Chen}}, \bibinfo {author} {\bibfnamefont {Y.}~\bibnamefont {Gao}}, \bibinfo
  {author} {\bibfnamefont {P.}~\bibnamefont {Wang}}, \bibinfo {author}
  {\bibfnamefont {Z.}~\bibnamefont {Mei}}, \bibinfo {author} {\bibfnamefont
  {Z.}~\bibnamefont {Pan}},\ and\ \bibinfo {author} {\bibnamefont {et~al.}},\
  }\bibfield  {title} {\bibinfo {title} {A comprehensive diagnostic system of
  ultra-thin liquid sheet targets},\ }\href
  {https://doi.org/10.1017/hpl.2023.101} {\bibfield  {journal} {\bibinfo
  {journal} {High Power Laser Science and Engineering}\ }\textbf {\bibinfo
  {volume} {12}},\ \bibinfo {pages} {e26} (\bibinfo {year} {2024})}\BibitemShut
  {NoStop}%
\bibitem [{\citenamefont {Zeil}\ \emph {et~al.}(2010)\citenamefont {Zeil},
  \citenamefont {Kraft}, \citenamefont {Bock}, \citenamefont {Bussmann},
  \citenamefont {Cowan}, \citenamefont {Kluge}, \citenamefont {Metzkes},
  \citenamefont {Richter}, \citenamefont {Sauerbrey},\ and\ \citenamefont
  {Schramm}}]{Zeil_2010}%
  \BibitemOpen
  \bibfield  {author} {\bibinfo {author} {\bibfnamefont {K.}~\bibnamefont
  {Zeil}}, \bibinfo {author} {\bibfnamefont {S.~D.}\ \bibnamefont {Kraft}},
  \bibinfo {author} {\bibfnamefont {S.}~\bibnamefont {Bock}}, \bibinfo {author}
  {\bibfnamefont {M.}~\bibnamefont {Bussmann}}, \bibinfo {author}
  {\bibfnamefont {T.~E.}\ \bibnamefont {Cowan}}, \bibinfo {author}
  {\bibfnamefont {T.}~\bibnamefont {Kluge}}, \bibinfo {author} {\bibfnamefont
  {J.}~\bibnamefont {Metzkes}}, \bibinfo {author} {\bibfnamefont
  {T.}~\bibnamefont {Richter}}, \bibinfo {author} {\bibfnamefont
  {R.}~\bibnamefont {Sauerbrey}},\ and\ \bibinfo {author} {\bibfnamefont
  {U.}~\bibnamefont {Schramm}},\ }\bibfield  {title} {\bibinfo {title} {The
  scaling of proton energies in ultrashort pulse laser plasma acceleration},\
  }\href {https://doi.org/10.1088/1367-2630/12/4/045015} {\bibfield  {journal}
  {\bibinfo  {journal} {New Journal of Physics}\ }\textbf {\bibinfo {volume}
  {12}},\ \bibinfo {pages} {045015} (\bibinfo {year} {2010})}\BibitemShut
  {NoStop}%
\bibitem [{\citenamefont {Higginson}\ \emph {et~al.}(2024)\citenamefont
  {Higginson}, \citenamefont {Lelièvre}, \citenamefont {Vassura},
  \citenamefont {Gugiu}, \citenamefont {Borghesi}, \citenamefont {Bernstein},
  \citenamefont {Bleuel}, \citenamefont {Goldblum}, \citenamefont {Green},
  \citenamefont {Hannachi} \emph {et~al.}}]{higginson2024}%
  \BibitemOpen
  \bibfield  {author} {\bibinfo {author} {\bibfnamefont {D.}~\bibnamefont
  {Higginson}}, \bibinfo {author} {\bibfnamefont {R.}~\bibnamefont
  {Lelièvre}}, \bibinfo {author} {\bibfnamefont {L.}~\bibnamefont {Vassura}},
  \bibinfo {author} {\bibfnamefont {M.}~\bibnamefont {Gugiu}}, \bibinfo
  {author} {\bibfnamefont {M.}~\bibnamefont {Borghesi}}, \bibinfo {author}
  {\bibfnamefont {L.}~\bibnamefont {Bernstein}}, \bibinfo {author}
  {\bibfnamefont {D.~L.}\ \bibnamefont {Bleuel}}, \bibinfo {author}
  {\bibfnamefont {B.~L.}\ \bibnamefont {Goldblum}}, \bibinfo {author}
  {\bibfnamefont {A.}~\bibnamefont {Green}}, \bibinfo {author} {\bibfnamefont
  {F.}~\bibnamefont {Hannachi}}, \emph {et~al.},\ }\bibfield  {title} {\bibinfo
  {title} {Global characterization of a laser-generated neutron source},\
  }\href {https://doi.org/10.1017/S0022377824000618} {\bibfield  {journal}
  {\bibinfo  {journal} {Journal of Plasma Physics}\ }\textbf {\bibinfo {volume}
  {90}},\ \bibinfo {pages} {905900308} (\bibinfo {year} {2024})}\BibitemShut
  {NoStop}%
\bibitem [{\citenamefont {Catford}(2019)}]{catkin2019}%
  \BibitemOpen
  \bibfield  {author} {\bibinfo {author} {\bibfnamefont {W.~N.}\ \bibnamefont
  {Catford}},\ }\href
  {https://personalpages.surrey.ac.uk/w.catford/kinematics/} {\bibinfo {title}
  {{Catkin: The Kinematics Programme in Excel}}} (\bibinfo {year} {2019}),\
  \bibinfo {note} {accessed: February 3, 2025}\BibitemShut {NoStop}%
\bibitem [{\citenamefont {Karsch}\ \emph {et~al.}(2003)\citenamefont {Karsch},
  \citenamefont {D\"usterer}, \citenamefont {Schwoerer}, \citenamefont {Ewald},
  \citenamefont {Habs}, \citenamefont {Hegelich}, \citenamefont {Pretzler},
  \citenamefont {Pukhov}, \citenamefont {Witte},\ and\ \citenamefont
  {Sauerbrey}}]{Karsch2003}%
  \BibitemOpen
  \bibfield  {author} {\bibinfo {author} {\bibfnamefont {S.}~\bibnamefont
  {Karsch}}, \bibinfo {author} {\bibfnamefont {S.}~\bibnamefont {D\"usterer}},
  \bibinfo {author} {\bibfnamefont {H.}~\bibnamefont {Schwoerer}}, \bibinfo
  {author} {\bibfnamefont {F.}~\bibnamefont {Ewald}}, \bibinfo {author}
  {\bibfnamefont {D.}~\bibnamefont {Habs}}, \bibinfo {author} {\bibfnamefont
  {M.}~\bibnamefont {Hegelich}}, \bibinfo {author} {\bibfnamefont
  {G.}~\bibnamefont {Pretzler}}, \bibinfo {author} {\bibfnamefont
  {A.}~\bibnamefont {Pukhov}}, \bibinfo {author} {\bibfnamefont
  {K.}~\bibnamefont {Witte}},\ and\ \bibinfo {author} {\bibfnamefont
  {R.}~\bibnamefont {Sauerbrey}},\ }\bibfield  {title} {\bibinfo {title}
  {High-intensity laser induced ion acceleration from heavy-water droplets},\
  }\href {https://doi.org/10.1103/PhysRevLett.91.015001} {\bibfield  {journal}
  {\bibinfo  {journal} {Phys. Rev. Lett.}\ }\textbf {\bibinfo {volume} {91}},\
  \bibinfo {pages} {015001} (\bibinfo {year} {2003})}\BibitemShut {NoStop}%
\bibitem [{\citenamefont {Brown}\ and\ \citenamefont
  {Jarmie}(1990)}]{Brown1990}%
  \BibitemOpen
  \bibfield  {author} {\bibinfo {author} {\bibfnamefont {R.~E.}\ \bibnamefont
  {Brown}}\ and\ \bibinfo {author} {\bibfnamefont {N.}~\bibnamefont {Jarmie}},\
  }\bibfield  {title} {\bibinfo {title} {Differential cross sections at low
  energies for $^{2}\mathrm{H}(d,p)^{3}\mathrm{H}$ and
  $^{2}\mathrm{H}(d,n)^{3}\mathrm{He}$},\ }\href
  {https://doi.org/10.1103/PhysRevC.41.1391} {\bibfield  {journal} {\bibinfo
  {journal} {Phys. Rev. C}\ }\textbf {\bibinfo {volume} {41}},\ \bibinfo
  {pages} {1391} (\bibinfo {year} {1990})}\BibitemShut {NoStop}%
\bibitem [{\citenamefont {Ying}\ \emph {et~al.}(1973)\citenamefont {Ying},
  \citenamefont {Cox}, \citenamefont {Barnes},\ and\ \citenamefont
  {Barrows}}]{Ying1973}%
  \BibitemOpen
  \bibfield  {author} {\bibinfo {author} {\bibfnamefont {N.}~\bibnamefont
  {Ying}}, \bibinfo {author} {\bibfnamefont {B.~B.}\ \bibnamefont {Cox}},
  \bibinfo {author} {\bibfnamefont {B.~K.}\ \bibnamefont {Barnes}},\ and\
  \bibinfo {author} {\bibfnamefont {A.~W.}\ \bibnamefont {Barrows}},\
  }\bibfield  {title} {\bibinfo {title} {A study of the
  $^{2}\mathrm{H}(d,p)^{3}\mathrm{H}$ and $^{2}\mathrm{H}(d,n)^{3}\mathrm{He}$
  reactions and the excited state of $^{4}\mathrm{He}$ at 23.9 {M}e{V}},\
  }\href {https://doi.org/10.1016/0375-9474(73)90080-8} {\bibfield  {journal}
  {\bibinfo  {journal} {Nucl. Phys. A}\ }\textbf {\bibinfo {volume} {206}},\
  \bibinfo {pages} {481} (\bibinfo {year} {1973})}\BibitemShut {NoStop}%
\bibitem [{\citenamefont {Kar}\ \emph {et~al.}(2016)\citenamefont {Kar},
  \citenamefont {Green}, \citenamefont {Ahmed}, \citenamefont {Alejo},
  \citenamefont {Robinson}, \citenamefont {Cerchez}, \citenamefont {Clarke},
  \citenamefont {Doria}, \citenamefont {Dorkings}, \citenamefont {Fernandez}
  \emph {et~al.}}]{Kar_2016}%
  \BibitemOpen
  \bibfield  {author} {\bibinfo {author} {\bibfnamefont {S.}~\bibnamefont
  {Kar}}, \bibinfo {author} {\bibfnamefont {A.}~\bibnamefont {Green}}, \bibinfo
  {author} {\bibfnamefont {H.}~\bibnamefont {Ahmed}}, \bibinfo {author}
  {\bibfnamefont {A.}~\bibnamefont {Alejo}}, \bibinfo {author} {\bibfnamefont
  {A.~P.~L.}\ \bibnamefont {Robinson}}, \bibinfo {author} {\bibfnamefont
  {M.}~\bibnamefont {Cerchez}}, \bibinfo {author} {\bibfnamefont
  {R.}~\bibnamefont {Clarke}}, \bibinfo {author} {\bibfnamefont
  {D.}~\bibnamefont {Doria}}, \bibinfo {author} {\bibfnamefont
  {S.}~\bibnamefont {Dorkings}}, \bibinfo {author} {\bibfnamefont
  {J.}~\bibnamefont {Fernandez}}, \emph {et~al.},\ }\bibfield  {title}
  {\bibinfo {title} {Beamed neutron emission driven by laser accelerated light
  ions},\ }\href {https://doi.org/10.1088/1367-2630/18/5/053002} {\bibfield
  {journal} {\bibinfo  {journal} {New Journal of Physics}\ }\textbf {\bibinfo
  {volume} {18}},\ \bibinfo {pages} {053002} (\bibinfo {year}
  {2016})}\BibitemShut {NoStop}%
\end{thebibliography}
